\documentclass[12pt]{article}

\usepackage[dvips]{color,graphicx}
\usepackage{psfrag}
\usepackage{amssymb}
\usepackage{amsmath}

\definecolor{gris}{rgb}{0.8,0.8,0.8}

\newcommand{\eq}[1]{\begin{equation} #1 \end{equation}}

\newcommand{\eqa}[1]{\begin{eqnarray} #1 \end{eqnarray}}
\newcommand{\al}[1]{\begin{align} #1 \end{align}}

\newcommand{\ds}{\displaystyle}

\newcommand{\nn}{\nonumber}

\newcommand{\n}{\noindent}

\newcommand{\azeL}{{A_0^L}}
\newcommand{\azeR}{{A_0^R}}
\newcommand{\azeLR}{{A_0^{L,R}}}
\newcommand{\apeL}{{A_\perp^L}}
\newcommand{\apeR}{{A_\perp^R}}
\newcommand{\apeLR}{{A_\perp^{L,R}}}
\newcommand{\apaL}{{A_\|^L}}
\newcommand{\apaR}{{A_\|^R}}
\newcommand{\apaLR}{{A_\|^{L,R}}}
\newcommand{\re}{{\rm Re}}
\newcommand{\im}{{\rm Im}}

\newcommand{\Ceff}[1]{{\cal C}^{\rm eff}_{#1}}
\newcommand{\Cpeff}[1]{{\cal C}^{\rm eff\prime}_{#1}}
\newcommand{\C}[1]{{\cal C}_{#1}}
\newcommand{\Cp}[1]{{\cal C}^{\prime}_{#1}}

\begin{document}

$\ $
\vspace{1.4cm}
\begin{center}
\Large\bf 
Complete Anatomy of $\bar B_d\to \bar K^{*0}(\to K\pi) \ell^+\ell^-$\\ and its angular distribution
\end{center}

\vspace{1mm}
\begin{center}
{\sc J.~Matias$\,^{a}$, F.~Mescia$\,^{b}$, M.~Ramon$\,^{a}$ and J.~Virto$\,^{a}$
}
\end{center}

\begin{center}
{\em 
$\,^{a}$ Universitat Aut\`onoma de Barcelona, 08193 Bellaterra, Barcelona, Spain\\[2mm]
$\,^{b}$ Departament d'Estructura i Constituents de la Mat\`eria and 
Institut de Ci\`encies del Cosmos (ICCUB), Universitat de Barcelona,
08028 Barcelona, Spain 
}
\end{center}

\vspace{1mm}
\begin{abstract}\noindent
\vspace{-5mm}

We present a complete and optimal set of observables for the exclusive 4-body $\bar B$ meson decay
$\bar B_d \to \bar K^{*0}(\to K\pi)\ell^+\ell^- $  in the low dilepton mass region,
 that contains a maximal number of clean observables. This basis of observables is built in a systematic way. We show that all the previously defined observables and any  observable that one can construct, can be expressed as a function of this basis. This set of observables contains all the information that can be extracted from the angular distribution in the cleanest possible way. We provide explicit expressions for the full and the uniangular distributions in terms of this basis. The conclusions presented here  can be easily extended to the large-$q^2$ region. We study the sensitivity of the observables to right-handed currents and scalars. Finally, we  present for the first time all the symmetries of the full distribution including massive terms and scalar contributions.

\end{abstract}


\section{Introduction}
\label{intro}

The rare decay $\bar B_d\to \bar K^{*0} (\to K\pi)\ell^+\ell^-$ provides unique opportunities in the search for New Physics in flavor physics due to the wealth and variety of angular observables available experimentally. A total of 12 different angular coefficients characterize its angular distribution, each being a function of the invariant squared mass of the lepton pair, $q^2$. Although a full angular analysis with a small $q^2$ binning requires a good deal of statistics, it constitutes a conceivable goal for LHCb, at least in its upgraded form.

First data on the decay rate and several angular observables are already available from BABAR, Belle, CDF and LHCb. BABAR \cite{babar} has measured the decay rate, the forward-backward asymmetry $A_{\rm FB}$ and the $K^*$ longitudinal polarization fraction $F_L$, all integrated separately in the low and high-$q^2$ regions. Concerning $q^2$-dependent measurements, Belle \cite{belle} has provided a measurement of the total branching ratio and $A_{\rm FB}$, while CDF \cite{cdf} has provided also measurements of $F_L$ as well as the observables $A_T^{(2)}$ and $A_{\rm im}$ (see \cite{kruger,matias1}), with a measurement of the $q^2$ dependence in the form of 3 bins in the low-$q^2$ region (below the $J/\psi$ resonance), a bin between the $J/\psi$ and $\psi '$ resonances, and two bins in the high-$q^2$ region (above the $\psi '$). The LHCb collaboration has also provided measurements of the branching ratio, $A_{\rm FB}$ and $F_L$ based on $\sim 300\ {\rm pb}^{-1}$ of data \cite{lhcb}, while a larger data set of 1 fb$^{-1}$ is on tape. In order to cope with limited statistics, near future plans focus on fully integrated observables, where the $q^2$ dependence is lost, at least within the low- and high-$q^2$ regions.

On the theoretical side, the interest is focused on the tayloring of observables with desired properties. These properties are: 1) a reduced hadronic uncertainty, and 2) an enhanced sensitivity to short distance contributions from New Physics (e.g. right handed currents, etc). Concerning hadronic uncertainties, the objective is to minimize the dependence on the soft form factors, which are difficult to compute and are the source of large theoretical uncertainties. This is achieved with the construction of ratios of angular observables where a complete LO cancellation of the form factors occurs. The search for observables with such desired properties has led to the formulation of a set of observables called $A_T^{(2)}$ \cite{kruger}, $A_T^{(3,4,5)}$ \cite{matias1,matias2} and $A_T^{\rm (re,im)}$ \cite{becirevic} at low-$q^2$, and an analogous set $H_T^{(1,2,3,4,5)}$ \cite{bobeth} at high-$q^2$. These observables have been studied in detail and they indeed exhibit a low theoretical uncertainty and a clean sensitivity to characteristic New Physics features.\footnote{For a representative set of references discussing the phenomenology of this decay mode see \cite{1,2,3,4,5,6,7,8,9,10,11}.}

The source of experimental input is the differential decay distribution  of the 4-body final state $\bar K^{*0}(\to K\pi) \ell^+\ell^-$ . It is described by four independent kinematic variables, which are traditionally chosen to be: the invariant squared mass $q^2$ of the lepton pair; the angle $\theta_K$ between the directions of flight of the kaon and the $\bar B$ meson in the rest frame of the $\bar K^{*0}$; the angle $\theta_l$ between the directions of flight of the $\ell^-$ and the $\bar B$ meson in the dilepton rest frame; and the azimutal angle $\phi$ between the two planes defined by the lepton pair and the $K\pi$ system\footnote{This definition of the kinematic variables coincides exactly with that of Refs.~\cite{bobeth,buras}}. In terms of these kinematic variables, the differential decay distribution can be written as
\eqa{\label{dist}
\frac{d^4\Gamma}{dq^2\,d\!\cos\theta_K\,d\!\cos\theta_l\,d\phi}&=&\frac9{32\pi} \bigg[
J_{1s} \sin^2\theta_K + J_{1c} \cos^2\theta_K + (J_{2s} \sin^2\theta_K + J_{2c} \cos^2\theta_K) \cos 2\theta_l\nn\\[1.5mm]
&&\hspace{-2.7cm}+ J_3 \sin^2\theta_K \sin^2\theta_l \cos 2\phi + J_4 \sin 2\theta_K \sin 2\theta_l \cos\phi  + J_5 \sin 2\theta_K \sin\theta_l \cos\phi \nn\\[1.5mm]
&&\hspace{-2.7cm}+ (J_{6s} \sin^2\theta_K +  {J_{6c} \cos^2\theta_K})  \cos\theta_l    
+ J_7 \sin 2\theta_K \sin\theta_l \sin\phi  + J_8 \sin 2\theta_K \sin 2\theta_l \sin\phi \nn\\[1.5mm]
&&\hspace{-2.7cm}+ J_9 \sin^2\theta_K \sin^2\theta_l \sin 2\phi \bigg]\,,
}
The explicit dependence of the coefficients $J_i(q^2)$ in terms of transversity amplitudes ($A_i$) is given in Section \ref{sec:2}.
The point to emphasize here is that only observables that respect the symmetries of this angular distribution can be obtained. These symmetries\footnote{Even if the term `symmetry" usually denotes  an invariance of the Lagrangian, in the present paper and following Ref.~\cite{matias2}, it refers to an invariance of the angular distribution under a rotation in the space of spin amplitudes.} are transformations among the transversity amplitudes that leave invariant the coefficients $J_i(q^2)$ of the angular distribution \cite{matias2}. The number of amplitudes and the number of such symmetries determine precisely the number of degrees of freedom that are available from an angular analysis alone. This number is given by \cite{matias2}
\eq{n_{\rm exp}=2\, n_A-n_s\ ,
\label{nexp}}
where $n_{\rm exp}$ denotes the number of experimental degrees of freedom, $n_A$ is the number of transversity amplitudes $A_j$ ($j=1,\dots,n_A$) and $n_s$ is the number of continuous transformations (or symmetries) of the $A$'s that leave the $J$'s invariant.

The number $n_s$ of continuous symmetries can be obtained by inspection of the set of infinitesimal transformations of the $A$'s, by the method described in Ref.~\cite{matias2}. In this way, one can infer the true number of independent experimental degrees of freedom $n_{\rm exp}$. If the number $n_J$ of coefficients $J_i$ is larger than $n_{\rm exp}$, then the $J_i$ are not independent observables, and a set of $(n_J-n_{\rm exp})$ relations must exist between them. As shown in Ref.~\cite{matias2}, there are exactly 8 independent $J$'s (and consequently 8 independent observables) in the case of massless leptons, and 10 considering the mass terms --not including the CP-conjugated mode--. Adding scalar contributions increases these numbers to 9 (with massless leptons)  and 12 including masses. The conclusion is that there is a definite number of independent experimental observables that can be extracted from the angular analysis, and this number $n_{\rm exp}$ is known from symmetry arguments. From a pragmatic standpoint, the symmetry formalism can be substituted by this set of $n_{\rm exp}$ independent observables\footnote{Still, we will use the symmetries at some points to furnish more formal proofs of certain aspects of the approach. Also, the symmetry formalism can be used, for instance, to  obtain explicit expressions of  the transversity amplitudes in terms of the coefficients $J_i$.}. This set of $n_{\rm exp}$ independent observables can be considered complete, in the sense that any additional observable can be expressed as a function of the observables in this set. Such a complete non-redundant set may be conveniently called a \emph{basis}.

In this context, the natural question is what is the best choice for the observables in the basis. The answer is not different from before: these observables should satisfy the desired properties of reduced hadronic uncertainty and good sensitivity to New Physics. The observables $A_T^{(2,3,4,5)}$, $A_T^{\rm (re, im)}$, $H_T^{(1,2,3,4,5)}$ proposed in Refs.~\cite{kruger, matias1,matias2,becirevic,bobeth} are excellent candidates, since they were designed to satisfy these requirements. The question is whether this set of observables contains a basis, or if other observables must be introduced.

The purpose of this paper is to give an answer to this issue and to construct, in a systematic way, an optimal basis of observables related to the angular distribution of the decay $\bar B_d\to \bar K^{*0}(\to K^+\pi^-)\ell^+\ell^-$, including lepton masses and scalar contributions (and valid in the presence of tensor operators). We do not consider in this paper the CP-conjugated mode nor the corresponding CP asymmetries. Anticipating some of the results of the paper, we shall see that:

\begin{enumerate}

\item The optimal basis contains two types of observables: observables with LO dependence on the soft form factors (Form Factor Dependent (FFD) observables) and observables free from this dependence at LO (Form Factor Independent (FFI) observables). The FFD observables suffer from large hadronic uncertainties due to the poorly known soft form factors. For this reason, the goal is to maximize the number of FFI observables in our basis. Examples of FFD observables are $J_i(q^2)$, $F_L$ or $A_{\rm im}$. We will choose as FFD observables for our basis the differential rate $d\Gamma/dq^2$ and the forward-backward asymmetry $A_{\rm FB}$, although other choices are also possible.

\item In the case of massless leptons, the optimal basis according  to the previous counting, must contain 8 independent observables: two FFD  observables ($d\Gamma/dq^2$ and $A_{\rm FB}$) and six clean FFI observables. Examples of FFI observables are $A_T^{(2,3,4,5)}$, $A_T^{\rm (re, im)}$ or $H_T^{(1,2,3,4,5)}$. FFI observables can be constructed in a systematic way as we will show in Section \ref{sec:3}, leading to a set of observables $P_{1,2,3,4,5,6}$, which we call \emph{primary observables}. $P_{1,2,3,4,5}$ are directly related to all already known observables, but $P_6$ is new, and it is necessary for obtaining full information from the angular distribution.

\item In the massive lepton case, the basis must contain 10 observables (2 FFD and 8 FFI). Eight of those observables are $d\Gamma/dq^2$, $A_{\rm FB}$ and $P_i$ with i=1...6. The two remaining FFI observables, which vanish in the massless limit, will be called  $M_1$ and $M_2$ and have never been considered before.

\item In the scalar case with massive leptons the counting of symmetries establishes the existence of 12 independent observables. The full set of observables is composed by the two FFD $d\Gamma/dq^2$ and $A_{\rm FB}$ together with 10 FFI $P_{1,2,3,4,5,6}$, $M_1$, $M_2$, $S_1$ and $S_2$. The two new observables $S_1$ and $S_2$ vanish in the absence of scalar contributions.

\item Any conceivable FFD observable can be written as a function of $d\Gamma/dq^2$, $A_{\rm FB}$, $P_i$,$M_i$ and $S_i$, but most importantly, any conceivable theoretically clean (FFI) observable can be written as a function of the $P_i$, $M_i$ and $S_i$ alone. At the same time, each of these observables contains unique information. It is in this sense that the basis is optimal and complete.

\end{enumerate}

The structure of the paper is the following. In Section \ref{sec:2} we briefly review the symmetry formalism for the angular distribution of $\bar B_d\to \bar K^{*0} (\to K\pi)\ell^+\ell^-$. In Section \ref{sec:3} we build the basis of observables in the massless case, and show how the full set of observables that have been considered in the literature can be recovered from this basis. We shall keep mass effects at this stage to make the generalization in Section \ref{sec:4} most straightforward. After presenting the generalization to the massive case, and introducing the massive observables, we write the full angular distribution in terms of the observables in the basis.

In Section \ref{sec:scalar} we include the effect of scalar operators, and show how the previous results are modified. In particular, we introduce two extra observables, $S_1$ and $S_2$, that vanish in the absence of scalar contributions. In Section \ref{uni} we present the most general expressions for the three uniangular distributions in terms of the observables in the basis.

In Section \ref{sensi} we study the New Physics sensitivity of the proposed observables. For that purpose we study the SM contribution including NLO effects using QCD Factorization, hadronic uncertainties and an estimate of $\Lambda/m_b$ corrections. We also consider how the SM expectations are modified in several NP scenarios. We analyze the position of the zeroes of the observables as well as those NP scenarios that affect most strongly each of the observables.

In  Section \ref{sec:sum} we summarize the relevant results of the paper. Finally, the core of the mathematical machinery related to the symmetry formalism, including constructive proofs of existence of the continuous symmetries has been collected in Appendix \ref{appA}. This appendix contains the explicit form of the symmetry transformations among the amplitudes in the massless (appendix \ref{appmassless}), the massive (appendix \ref{appmassive}) and scalar case (appendix \ref{appscalar}). In appendix \ref{largerec} we present the building blocks  of the observables in the large recoil limit.

\section{Symmetries of the angular distribution}
\label{sec:2}

The coefficients of the distribution given in Eq.~(\ref{dist}) can be written in terms of transversity amplitudes. In the massless case there are six such complex amplitudes: $A_0^{R,L}$, $A_\|^{R,L}$ and $A_\perp^{R,L}$. An additional complex amplitude $A_t$ is required in the massive case, and in the presence of scalar contributions a new amplitude $A_S$ must be included. The expressions for these coefficients read,
\eqa{
J_{1s}  & = & \frac{(2+\beta_\ell^2)}{4} \left[|\apeL|^2 + |\apaL|^2 +|\apeR|^2 + |\apaR|^2 \right]
    + \frac{4 m_\ell^2}{q^2} \re\left(\apeL\apeR^* + \apaL\apaR^*\right)\,,\nn\\[1mm]
J_{1c}  & = &  |\azeL|^2 +|\azeR|^2  + \frac{4m_\ell^2}{q^2} \left[|A_t|^2 + 2\re(\azeL^{}\azeR^*) \right] + \beta_\ell^2\, |A_S|^2 \,,\nn\\[1mm]
J_{2s} & = & \frac{ \beta_\ell^2}{4}\left[ |\apeL|^2+ |\apaL|^2 + |\apeR|^2+ |\apaR|^2\right],
\hspace{0.92cm}    J_{2c}  = - \beta_\ell^2\left[|\azeL|^2 + |\azeR|^2 \right]\,,\nn\\[1mm]
J_3 & = & \frac{1}{2}\beta_\ell^2\left[ |\apeL|^2 - |\apaL|^2  + |\apeR|^2 - |\apaR|^2\right],
\qquad   J_4  = \frac{1}{\sqrt{2}}\beta_\ell^2\left[\re (\azeL\apaL^* + \azeR\apaR^* )\right],\nn \\[1mm]
J_5 & = & \sqrt{2}\beta_\ell\,\Big[\re(\azeL\apeL^* - \azeR\apeR^* ) - \frac{m_\ell}{\sqrt{q^2}}\,
\re(\apaL A_S^*+ \apaR^* A_S) \Big]\,,\nn\\[1mm]
J_{6s} & = &  2\beta_\ell\left[\re (\apaL\apeL^* - \apaR\apeR^*) \right]\,,
\hspace{2.25cm} J_{6c} = 4\beta_\ell\, \frac{m_\ell}{\sqrt{q^2}}\, \re (\azeL A_S^*+ \azeR^* A_S)\,,\nn\\[1mm]
J_7 & = & \sqrt{2} \beta_\ell\, \Big[\im (\azeL\apaL^* - \azeR\apaR^* ) +
\frac{m_\ell}{\sqrt{q^2}}\, \im (\apeL A_S^* - \apeR^* A_S)) \Big]\,,\nn\\[1mm]
J_8 & = & \frac{1}{\sqrt{2}}\beta_\ell^2\left[\im(\azeL\apeL^* + \azeR\apeR^*)\right]\,,
\hspace{1.9cm} J_9 = \beta_\ell^2\left[\im (\apaL^{*}\apeL + \apaR^{*}\apeR)\right] \,,
\label{Js}}\\
where the parameter $\beta_\ell$ is given by
\eq{\beta_\ell=\sqrt{1-\frac{4m_\ell^2}{q^2}}\ .
\label{beta}}
We will distinguish between the three cases of interest: massless leptons and no scalar amplitude, massive leptons and no scalar amplitude, and massive leptons plus a scalar amplitude. We also show that tensor contributions cannot change the picture.\\

\n {\bf A. Massless leptons, no scalars}\\[2mm]
In this case we can put $A_S\to 0$, drop the $m_\ell^2$ terms and set $\beta_\ell\to 1$. There are in total six complex transversity amplitudes: $A_0^{R,L}$, $A_\|^{R,L}$ and $A_\perp^{R,L}$,  which add up to $2n_A=12$ real theoretical quantities. However, an infinitesimal transformation of the distribution (see \cite{matias2}) shows that there must be $n_s=4$ continuous transformations between the $A_i^{L,R}$ that leave invariant the angular distribution. Two of them are simple phase transformations: $A_i^{L}\to e^{i\phi_L}A_i^{L}$ and $A_i^{R}\to e^{i\phi_R}A_i^{R}$, while the other two mix $L$ and $R$ amplitudes (see Appendix \ref{appA} and Ref.~\cite{matias2} for the explicit form of these transformations). According to Eq.~(\ref{nexp}), there must be precisely $n_{\rm exp}=8$ independent observables. This implies, in turn, that there should be 4 relationships among the 12 coefficients $J_i(q^2)$. Three of them are straightforward: $J_{6c}=0$, $J_{1s}=3J_{2s}$ and $J_{1c}=-J_{2c}$, while the remaining one is more involved \cite{matias2}:
\eqa{
J_{2c} &=& - 2\,\frac{ (2 J_{2s}+ J_3) \left(4 J_4^2+\beta_\ell^2 J_7^2\right) + ( 2 J_{2s} - J_3) \left(\beta_\ell^2 J_5^2+4 J_8^2 \right)}{16 J_{2s}^{2} -  \left(4 J_3^2+ \beta_\ell^2 J_{6s}^{2} + 4 J_9^2 \right)}\nn\\
&&+ 4\, \frac{\beta_\ell^2 J_{6s} (J_4 J_5 + J_7 J_8) + J_9 (\beta_\ell^2 J_5 J_7 - 4 J_4 J_8)}{16 J_{2s}^{2} -  \left(4 J_3^2+ \beta_\ell^2 J_{6s}^{2} + 4 J_9^2 \right)}\ ,
\label{relation}}
where $\beta_\ell \to 1$ should be understood in the massless case. The derivation of this expression requires the symmetry formalism and will be outlined later.\\

\n {\bf B. Massive leptons, no scalars}\\[2mm]
In this case we just set $A_S\to 0$ in Eq.~(\ref{Js}). Now there are seven complex transversity amplitudes, including $A_t$, which add up to $2n_A=14$ real theoretical quantities. As discussed in Appendix \ref{appmassive}, there are $n_s=4$ continuous transformations that leave the $J_i$ invariant (these are different from the symmetries of the massless case, since those are broken by mass effects). Two of these symmetries are phase rotations: $A_t\to e^{i\phi_t} A_t$ and $A_{0,\|,\bot}^{L,R}\to e^{i\phi} A_{0,\|,\bot}^{L,R}$, while the other two are nonlinear transformations (see Appendix \ref{appmassive}). According to Eq.~(\ref{nexp}), there must be precisely $n_{\rm exp}=10$ independent observables, which means that 2 relationships between the coefficients $J_i$ can be found. The relationships $J_{1s}=3J_{2s}$ and $J_{1c}=-J_{2c}$ are no longer satisfied; however, $J_{6c}=0$ and Eq.~(\ref{relation}) remain exactly true in the massive case (this was the reason for keeping the factors $\beta_\ell$ explicit in Eq.~(\ref{relation})).
Notice that in Ref.~\cite{matias2} the discussion was limited to the massless lepton case, so Eq.~(\ref{relation})  generalizes the relation in Ref.~\cite{matias2} to the massive case.\\

\n {\bf C. Massive leptons plus scalars}\\[2mm]
In this case we deal with 8 complex transversity amplitudes, which add up to $2n_A=16$ real theoretical variables. As demonstrated in Appendix \ref{appscalar} (see also Ref.~\cite{matias2}), there are $n_s=4$ symmetries. This means there must be exactly $n_{\rm exp}=12$ independent observables, which implies that all the $J_i$ are independent, and none of the previous relations hold in this case.\\

\n {\bf D. Massive leptons, scalars and tensors}\\[2mm]
The fact that all the $J_i$ are independent in the scalar case can be used to go a bit further in the reasoning. Imagine we wanted to include NP contributions from tensor operators. Then we would expect, at least, a new amplitude $A_T$ modifying somehow the angular distribution [Eqs.~(\ref{Js})]. However, according to Eq.~(\ref{nexp}) and since $n_{\rm exp}$ is as large as it can be, for each new amplitude $A_T$ there must be two extra symmetries. These symmetries must disappear in the limit $A_T\to 0$, and therefore they can be used to set $A_T\to 0$. Another way to see this is the following: since the $J_i$ are all independent in the scalar case, one can always obtain a set of $A^{R,L\,\prime}_{\|,\bot,0}$, $A_t'$, $A_S'$ that reproduce the angular distribution in the presence of new amplitudes such as $A_T$. Therefore, new tensor operators can only give new contributions to existing amplitudes, meaning that the basis of observables defined in the scalar case remains unchanged in the presence of tensors.\\

We will now discuss in turn the relevant set of observables that one can consider in each of these three cases of interest.

\section{Observables for  massless leptons}
\label{sec:3}

Not any observable constructed from the transversity amplitudes can be obtained from the angular distribution. As a necessary and sufficient condition, such an observable must be invariant under the symmetry transformations of the transversity amplitudes $A$'s; we then say that the observable respects the symmetries of the angular distribution.  Fortunately, there exists a systematic procedure to construct all such possible observables. 

We start defining the following complex vectors \cite{matias2},
\eq{
n_\|=\binom{A_\|^L}{A_\|^{R*}}\ ,\quad
n_\bot=\binom{A_\bot^L}{-A_\bot^{R*}}\ ,\quad
n_0=\binom{A_0^L}{A_0^{R*}}\ .
}
With these vectors we can construct the products $|n_i|^2= n_i^\dagger n_i$ and $n_i^\dagger\, n_j$,
\eq{
\begin{array}{rclrcl}
|n_\||^2&=&|A_\|^L|^2+|A_\|^{R}|^2= \dfrac{2 J_{2s} -J_3}{\beta_\ell^2}\ , \quad\! &
n_\bot^\dagger\, n_\| &=& A_\bot^{L*} A_\|^L-A_\bot^{R} A_\|^{R*}= \dfrac{\beta_\ell J_{6s} -2 i J_9}{2{\beta_\ell^2}}\ ,\\[5mm]
|n_\bot|^2&=&|A_\bot^L|^2+|A_\bot^{R}|^2=\dfrac{2 J_{2s} +J_3}{\beta_\ell^2}\ , \quad\! &
n_0^\dagger\, n_\|&=& A_0^{L*} A_\|^{L}+A_0^{R} A_\|^{R*}= \dfrac{2 J_{4} - i \beta_\ell J_7}{\sqrt{2}\beta_\ell^2}\ ,\\[5mm]
|n_0|^2&=&|A_0^L|^2+|A_0^{R}|^2=-\dfrac{J_{2c}}{\beta_\ell^2}\ , \quad\! &
n_0^\dagger\, n_\bot&=& A_0^{L*} A_\bot^{L}-A_0^{R} A_\bot^{R*}=  \dfrac{\beta_\ell J_5 - 2i J_8}{\sqrt{2}\beta_\ell^2}\ .
\end{array}
\label{ns}}
These quantities automatically respect the symmetries of the angular distribution, since they can be expressed in terms of the $J_i$. Considering real and imaginary parts, there are 9 real quantities that encode all the information of the angular distribution, and by combining them one can construct systematically all possible allowed observables consistent with the symmetry requirements. However they are not all independent: any set of complex 2-vectors $\{n_0$, $n_\|$, $n_\bot\}$ satisfies
\eq{\big|(n_\|^\dagger\, n_\bot) |n_0|^2 - (n_\|^\dagger\, n_0)(n_0^\dagger\, n_\bot)\big|^2
= (|n_0|^2 |n_\||^2 - |n_0^\dagger\, n_\||^2)  (|n_0|^2 |n_\bot|^2 - |n_0^\dagger\, n_\bot|^2)\ .}
Using Eqs.~(\ref{ns}), this relation translates precisely into the relation for the $J_i$ given in Eq.~(\ref{relation}).

Now that the formalism assures the systematic construction of observables that respect the symmetries of the angular distribution, we must focus on the cancellation of hadronic form factors. At leading order in $1/m_b$ and $\alpha_s$, and at large recoil ($E_{K^*} \to \infty$), the transversity amplitudes $\azeLR$, $\apaLR$ and $\apeLR$ can be written as:
\eqa{
\apeLR &=&\sqrt{2} N m_B(1- \hat s)\bigg[  (\Ceff9 + \Cpeff9) \mp (\C{10} + \Cp{10})
+\frac{2\hat{m}_b}{\hat s} (\Ceff7 + \Cpeff7) \bigg]\xi_{\bot}(E_{K^*})  \nn \\[2mm]
\apaLR &=& -\sqrt{2} N m_B (1-\hat s)\bigg[(\Ceff9 - \Cpeff9) \mp (\C{10} - \Cp{10}) 
+\frac{2\hat{m}_b}{\hat s}(\Ceff7 - \Cpeff7) \bigg] \xi_{\bot}(E_{K^*}) \nn \\[2mm]
\azeLR  &=& -\frac{N m_B (1-\hat s)^2}{2 \hat{m}_{K^*} \sqrt{\hat s}} \bigg[ (\Ceff9 - \Cpeff9)  \mp (\C{10} - \Cp{10}) + 2\hat{m}_b (\Ceff7 - \Cpeff7) \bigg]\xi_{\|}(E_{K^*})
\label{LargeRecoilAs}}
where $\hat s = q^2 /m_B^2$, $\hat{m}_i = m_i/m_B$, and terms of $O(\hat{m}_{K^*}^2)$ have been neglected. The normalization is given by
\eq{N=V_{tb}V_{ts}^*\sqrt{\frac{\beta_\ell G_F^2 \alpha^2 q^2 \lambda^{1/2}}{3\cdot 2^{10} \pi^5 m_B^3}}\ ,}
with $\lambda=[q^2-(m_B+m_{K^*})^2][q^2-(m_B-m_{K^*})^2]$.
Therefore, at first order, we have $n_0\propto \xi_\|$ and $n_\|,n_\bot \propto \xi_\bot$. This establishes a clear guideline in the construction of clean observables, as ratios of quantities in Eq.~(\ref{ns}) where the $\xi_{\|,\bot}$ cancel [Form Factor Independent (FFI) observables].

Before providing a complete list of observables constructed according to this procedure, we should note the following. There are 8 independent quantities in Eq.~(\ref{ns}) that constitute the building blocks of the observables. The soft form factors $\xi_{\|,\bot}$ can be thought of as 2 irreducible normalization factors in the products $n_i^\dagger n_j$, and therefore one cannot construct 8 independent combinations where the soft form factors cancel. The best one can do is to construct 6 clean observables, plus 2 observables that contain the information on the two form factors ---or Form Factor Dependent (FFD) observables--.

For these two FFD observables we can choose, quite naturally, the angular-integrated differential decay rate $d\Gamma/dq^2$, and the forward-backward asymmetry $A_{\rm FB}$:
\eqa{
\frac{d\Gamma}{dq^2} &=& \int d\!\cos\theta_l\, d\!\cos\theta_K d\phi\,
\frac{d^4\Gamma}{dq^2\,d\!\cos\theta_K\,d\!\cos\theta_l\,d\phi}
=\dfrac{1}{4} \left(3 J_{1c} + 6 J_{1s} - J_{2c} -2 J_{2s}\right)\ ,\hspace{0.8cm}\label{dgamma}\\[2mm]
A_{\rm FB} &=& \frac{1}{d\Gamma/dq^2} \left[ \int_{-1}^0 - \int_0^1 \right]d\!\cos\theta_l\, \frac{d^2\Gamma}{dq^2 d\!\cos\theta_l} = -\dfrac{3 J_{6s}}{3J_{1c}+6J_{1s}-J_{2c}-2J_{2s}}\ .
\label{AFB}
}
Notice that, while Eq.~(\ref{dgamma}) is completely general, in the last equality of Eq.~(\ref{AFB}) we have assumed that $J_{6c}=0$ due to the absence of scalar contributions. In the massless case, since $J_{1s}=3J_{2s}$ and $J_{1c}=-J_{2c}$, these expressions simplify to $d\Gamma/dq^2=J_{1c}+4J_{2s}$ and $A_{\rm FB}=-3 J_{6s}/[4(J_{1c}+4J_{2s})]$.

For the six (clean) FFI observables we choose the following set:
\eqa{
P_1&=&\frac{|n_\bot|^2-|n_\||^2}{|n_\bot|^2+|n_\||^2}=\frac{ J_{3}}{2 J_{2s}}\ ,\label{p1}\\
P_2&=&\frac{{\rm Re}(n_\bot^\dagger\, n_\|)}{|n_\||^2+|n_\bot|^2}=\beta_\ell\frac{J_{6s}}{8 J_{2s}}\ ,\label{p2}\\
P_3&=&\frac{{\rm Im}(n_\bot^\dagger\, n_\|)}{|n_\||^2+|n_\bot|^2}=-\frac{J_{9}}{4 J_{2s}}\ ,\label{p3}\\
P_4&=&\frac{{\rm Re}(n_0^\dagger\, n_\|)}{\sqrt{|n_\||^2 |n_0|^2}}
=\frac{\sqrt{2}J_4}{\sqrt{-J_{2c}(2 J_{2s}-J_{3})}}\ ,\\
P_5&=&\frac{{\rm Re}(n_0^\dagger\, n_\bot)}{\sqrt{|n_\bot|^2 |n_0|^2}}
=\dfrac{\beta_\ell J_5}{\sqrt{-2 J_{2c}(2 J_{2s}+J_3)}}\ ,\label{p5}\\
P_6&=&\frac{{\rm Im}(n_0^\dagger\, n_\|)}{\sqrt{|n_\||^2 |n_0|^2}}
=-\frac{\beta_\ell J_7}{\sqrt{-2 J_{2c}(2 J_{2s}-J_3)}}\ ,\label{p6}
\label{props}}
although other similar ratios are possible. We have used the following criteria for choosing among the different possible FFI observables: (1) they are simple ratios of the quantities in Eq.~(\ref{ns}) where the form factors $\xi_{\bot,\|}$ cancel, (2) they take values in the range [-1,1], and (3) they show good sensitivity to selected New Physics (see Section \ref{sensi}).
To summarize, the complete basis of observables in the massless case is given by:
\eq{O_{m_\ell=0}=\Big\{ \frac{d\Gamma}{dq^2}, A_{\rm FB}, P_1, P_2, P_3, P_4, P_5, P_6  \Big\} }
where the six $P_i$ are clean observables.

As mentioned, all possible observables can be expressed in terms of the observables in the basis. In particular, all known observables can be related to this set. For example, the usual FFI observables can be expressed in terms of $P_{1,2,3,4,5,6}$:
\eq{
\begin{array}{rclrcl}
A_T^{(2)}&=& P_1\ ,\qquad & A_T^{(5)}&=& \frac12 \sqrt{1-P_1^2-4 P_2^2-4 P_3^2}\ ,\\[2mm]
A_T^{\rm (re)} &=& 2 P_2 \ ,\qquad & A_T^{\rm (im)}&=& -2 P_3\ ,\\[2mm]
H_T^{(1)}&=&P_4 \ ,\qquad & H_T^{(2)}&= & P_5.
\end{array}
}
Also, the relationship $(2 A_T^{(5)})^2+(A_T^{(2)})^2+(A_T^{\rm (re)})^2+(A_T^{\rm (im)})^2=1$, presented in Ref.~\cite{becirevic}, follows trivially in terms of the $P_i$.

In the case of the observables $A_T^{(3)}$ and $A_T^{(4)}$, the corresponding expressions are more involved. They can be expressed in terms of our basis by first recovering their expression in terms of the $J$'s:
\begin{equation}
A_T^{(3)}= \sqrt{\frac{4  J_4^2 + \beta_\ell^2 J_7^2}{-2 J_{2c} (2 J_{2s}+J_3)}}\ , \quad \quad 
A_T^{(4)} =  \sqrt{\frac{4  J_8^2 + \beta_\ell^2 J_5^2}{4 J_4^2 + \beta_\ell^2 J_7^2 }}\ ,
\end{equation}
and substituting the $J$'s in terms of our basis of observables as given in Eqs.~(\ref{J1s})-(\ref{J9}) in the following section.

The same can be done for all the known FFD observables, which can be expressed in terms of $P_i$, $d\Gamma/dq^2$ and $A_{\rm FB}$, for example:
\eqa{
F_T&=&1-F_L= -\frac{2\beta_\ell}{3}\frac{A_{\rm FB}}{P_2} \label{FT}\\
A_{\rm im} &=& \frac{2}{3} \frac{A_{\rm FB} P_3}{P_2} \label{Aim}
}

\section{Observables in the massive case}
\label{sec:4}

All the coefficients in the massive case can be expressed in terms of the quantities of Eq.~(\ref{ns}), with the exception of $J_{1c}$ and $J_{1s}$, which can be written as:
\eqa{
J_{1s}&=& \frac{2+\beta_\ell}{4}\,\big[|n_\bot|^2 + |n_\||^2\big]
+ \frac{2m_\ell^2}{q^2}\,\big[n_\|^T\sigma_1
n_\| + n_\|^\dagger\sigma_1 n_\|^* - n_\bot^T\sigma_1 n_\bot - n_\bot^\dagger\sigma_1 n_\bot^*\big]\ ,\\
J_{1c}&=& |n_0|^2 + \frac{2m_\ell^2}{q^2}\,
\big[2|A_t|^2 + n_0^T\sigma_1 n_0 + n_0^\dagger \sigma_1 n_0^*\big]\ .
}
where $\sigma_1$ is the Pauli matrix:
\eq{\sigma_1=\left(\begin{array}{cc} 0 & 1\\1&0 \end{array}\right).}
The important point demonstrated in Appendix \ref{appmassive} is that in this case the symmetries, like in the massless case, can be expressed as a single unitary rotation $U$ on $n_{0,\bot,\|}$. This means that all the products [Eq.(\ref{ns})] that were invariant in the massless case are still invariant under the new symmetries (so all the observables defined in Section \ref{sec:3} are still valid), which in turn means that the new terms appearing in $J_{1c}$ and $J_{1s}$: 

$$  \frac{4 m_\ell^2}{q^2} \re\left(\apeL\apeR^* + \apaL\apaR^*\right)\ ,  \quad \quad \frac{4m_\ell^2}{q^2} \left[|A_t|^2 + 2\re(\azeL^{}\azeR^*) \right]\ ,
$$
must be invariant by themselves. This is a key idea in order to find the continuous symmetries (see Appendix \ref{appmassive}), but it is also crucial to the construction of the new observables, the obvious ones being:
\eqa{
M_1&=&  \frac{2m_\ell^2}{q^2}\frac1{\beta_\ell^2}\cdot\frac{n_\|^T\sigma_1
n_\| + n_\|^\dagger\sigma_1 n_\|^* - n_\bot^T\sigma_1 n_\bot - n_\bot^\dagger\sigma_1 n_\bot^*}{|n_\||^2+|n_\bot|^2}=\frac{\beta_\ell^2 J_{1s}-(2+\beta_\ell^2)J_{2s}}{4 \beta_\ell^2 J_{2s}}\ , \label{m1}\\
M_2&=& \frac{2m_\ell^2}{q^2}\cdot \frac{2|A_t|^2 + n_0^T\sigma_1 n_0 + n_0^\dagger
\sigma_1 n_0^*}{|n_0|^2}=-\frac{\beta_\ell^2 J_{1c}+J_{2c}}{J_{2c}}\ .\label{m2}
}
We note that these observables are of the FFI type, and thus theoretically clean. This can be inferred from the large recoil limit expressions (\ref{LargeRecoilAs}) and
\eq{A_t=\frac{N m_B}{2\hat m_{K^*}\sqrt{\hat s}}(1-\hat s)^2 \big[2 (\C{10}-\Cp{10}) + \frac{q^2}{m_\ell} (\C{P} - \C{P}^\prime) \big]\, \xi_\|(E_{K^*})\ .
\label{At}}
Moreover, $M_1$ and $M_2$ vanish in the massless limit (from the right hand side of Eqs.~(\ref{m1}) and (\ref{m2}) one can see that this follows from the relationships of the massless case, $J_{1s}=3J_{2s}$ and $J_{1c}=-J_{2c}$, that are broken for non-zero lepton masses; in fact $M_1$ measures the breaking of the relation $J_{1s}=3J_{2s}$, while $M_2$ measures the breaking of $J_{1c}=-J_{2c}$).

From the discussion in Section \ref{sec:2}, together with the observations that $M_1$ and $M_2$ vanish for $m_\ell\to 0$, and that the observables of the massless case are still valid, one concludes that these two observables complete the basis of 10 independent observables of the massive case. This basis is:
\eq{O_{m_\ell\ne 0}=\Big\{ \frac{d\Gamma}{dq^2}, A_{\rm FB}, P_1, P_2, P_3, P_4, P_5, P_6,M_1,M_2  \Big\}\ . }
The coefficients $J_i$ of the angular distribution are themselves observables (of the FFD type), and it is interesting to express them in terms of $d\Gamma/dq^2$, $A_{\rm FB}$ and $P_i$. The explicit expressions read:

\al{
J_{1s} &= -  \dfrac{2+(1+4 M_1)\beta_\ell^2}{6 \beta_\ell} \chi\frac{d\Gamma}{dq^2}\ , \label{J1s} \\
J_{1c} &= \dfrac{2 (1+M_2)}{3 \beta_\ell (3+3 M_2 + \beta_\ell^2)} \left( 6 \beta_\ell + [3+(1+6 M_1) \beta_\ell^2]\, \chi\right) \frac{d\Gamma}{dq^2}\ ,\\
J_{2s} &= -\dfrac{\beta_\ell}{6} \chi\frac{d\Gamma}{dq^2}\ , \label{J2s}}
\al{
J_{2c} &= -\dfrac{2 \beta_\ell}{3(3+3 M_2 + \beta_\ell^2)}  \left( 6 \beta_\ell + [3+(1+6 M_1) \beta_\ell^2]\, \chi\right) \frac{d\Gamma}{dq^2}\ ,\\
J_3  &= -\dfrac{\beta_\ell P_1}{3} \chi\frac{d\Gamma}{dq^2}\ ,\\
J_4 &= \dfrac{P_4}{3} \sqrt{\dfrac{(P_1-1) \beta_\ell^2}{3+3 M_2 + \beta_\ell^2}  \left( 6 \beta_\ell + [3+(1+6 M_1) \beta_\ell^2]\, \chi\right)\chi}\ \frac{d\Gamma}{dq^2}\ ,\\
J_5 &= \dfrac{2 P_5}{3} \sqrt{\dfrac{(-P_1-1)}{3+3 M_2 + \beta_\ell^2}  \left( 6 \beta_\ell + [3+(1+6 M_1) \beta_\ell^2]\, \chi\right)\chi}\ \frac{d\Gamma}{dq^2}\ ,\\
J_{6s} &= -\dfrac{4 P_2}{3} \chi \frac{d\Gamma}{dq^2}\ ,\\
J_7 &=-\dfrac{2 P_6}{3} \sqrt{\dfrac{(P_1-1)}{3+3 M_2 + \beta_\ell^2}  \left( 6 \beta_\ell + [3+(1+6 M_1) \beta_\ell^2]\, \chi\right)\chi}\ \frac{d\Gamma}{dq^2}\ ,\\
J_9 &= \dfrac{2 \beta_\ell P_3}{3} \chi\frac{d\Gamma}{dq^2}\ ,  \label{J9}  }
where, in this case,
\eq{\chi=\frac{A_{\rm FB}}{P_2}\ .}
The observable $\chi$ is well defined because $A_{\rm FB}$ and $P_2$ share the same zeroes [see Eqs.~(\ref{AFB}) and (\ref{p2})]. This will change after including scalars, but also $\chi$ will change. On the other hand, since $J_{1c}>J_{2c}$  [as can be checked from Eq.~(\ref{Js})], it follows that $3+3 M_2 + \beta_\ell^2>0$, and no vanishing denominators can occur in Eqs.~(\ref{J1s})-(\ref{J9}).

Notice that $J_8$ is absent from this list, as it is not an independent coefficient in the absence of scalar contributions. In order to obtain its expression in terms of the observables one should write $J_8$ as a function of the other $J_i$ using Eq.~(\ref{relation}), and then plug in the expressions (\ref{J1s})-(\ref{J9}). Alternatively, one might want to leave $J_8$ as a free parameter when fitting the angular distribution in terms of the observables, and then check the relationship  in Eq.~(\ref{relation}), to look for scalar contributions (see next section).
 
Using the substitutions $M_1,M_2 \rightarrow 0$ and $\beta_\ell \rightarrow 1$, Eqs.~(\ref{J1s})-(\ref{J9}) transform into the corresponding massless case expressions.

\section{Inclusion of Scalar Operators}
\label{sec:scalar}

In the presence of scalar operators, a new amplitude $A_S$ appears. This makes 16 the number of real theoretical degrees of freedom (8 complex amplitudes): $2n_A=16$. There are 4 symmetries, making $n_{\rm exp}=12$. This means that all the $J_i$ are independent, which in turn implies that the relationship $J_{6c}=0$ and Eq.~(\ref{relation}) do not hold any longer. This observation allows us to choose the two extra observables that are needed in the basis in the presence of scalar contributions, as the amount by which these relationships are broken. This provides two independent observables, $S_1$ and $S_2$, that vanish in the absence of scalars. The first observable is:
\eq{
S_1 = -\frac{\beta_\ell\sqrt{q^2}}{4 m_\ell}\frac{J_{6c}}{J_{2c}}\ ,
\label{s1}}
which measures the breaking of the relation $J_{6c}=0$. The second one is:

\eqa{
S_2 &=& \frac{16 J_{2c}J_{2s}^2 - 4 J_{2c} J_3^2 + 16 J_{2s} J_4^2 + 8 J_3 J_4^2 + 16 J_{2s} J_8^2 - 8 J_3 J_8^2 + 16 J_4 J_8 J_9 - 4 J_{2c} J_9^2}{J_{2c} J_{2s}^2}\nn\\
&&\hspace{-1cm} +\ \beta^2  \frac{4 J_{2s} J_5^2 - 2 J_3 J_5^2 - 4 J_4 J_5 J_{6s} - J_{2c} J_{6s}^2 + 4 J_{2s} J_7^2 + 2 J_3 J_7^2 - 4 J_{6s} J_7 J_8 - 4 J_5 J_7 J_9}{J_{2c} J_{2s}^2}\ ,\qquad
\label{s2}}
which gives a measure of the violation of Eq.~(\ref{relation}). One can easily check that both observables are of the FFI type, by noting that the large recoil expression for $A_S$ is:
\eq{A_S=-\,\frac{N m_B^2}{\hat m_{K^*}}(1-\hat s)^2 \big[ \C{S} - \C{S}^\prime \big]\, \xi_\|(E_{K^*})\ .
\label{As}}
While most of the results in the previous sections remain unchanged in the presence of scalars, some differences must be clarified. When $A_S\ne 0$, the observables $P_5$, $P_6$ and $M_2$ get modified. In particular Eqs.~(\ref{p5}), (\ref{p6}) and (\ref{m2}) in terms of the vectors $n_i$ do not hold since new terms proportional to $A_S$ arise, and these observables must be redefined. The simplest way to generalize these observables in presence of scalars is simply  to use their definition  in terms of the $J_i$ in Eqs.~(\ref{p5}), (\ref{p6}) and (\ref{m2}).
These three observables are, together with $S_1$ and $S_2$, the only ones sensitive to scalar contributions. With this in mind, an optimal basis of observables in the presence of scalars reads:
\eq{O_{m_\ell\ne 0}^{\rm scalars}=\Big\{ \frac{d\Gamma}{dq^2}, A_{\rm FB}, P_1, P_2, P_3, P_4, P_5, P_6,M_1,M_2,S_1,S_2  \Big\}\ . }
In the case of $A_{\rm FB}$, we keep the definition in Eq.~(\ref{AFB}) as the angular integral, which means that now:
\eq{A_{FB}= -\dfrac{3 J_{6s}+3 J_{6c}/2}{3J_{1c}+6J_{1s}-J_{2c}-2J_{2s}}\ .
\label{AFBsc}}
Because of this, Eqs.~(\ref{J1s})-(\ref{J9}) are modified slightly. First, when $S_i\ne 0$ we have:
\eq{\chi = \frac{(3+3M_2+\beta_\ell^2)A_{\rm FB} + (m_\ell/\sqrt{q^2})\, 6\beta_\ell S_1}{(3+3M_2+\beta_\ell^2)P_2 - (m_\ell/\sqrt{q^2}) (3+6\beta_\ell^2 M_1+\beta_\ell^2) S_1}
\label{P2bar}}
in (\ref{J1s})-(\ref{J9}). Furthermore, $J_{6c}$ is non-zero:
\eq{
J_{6c} = \frac{8 m_\ell S_1}{3\sqrt{q^2}}\frac{(6\beta_\ell^2 M_1+\beta_\ell^2 +3) A_{\rm FB} + 6\beta_\ell P_2}{(3 M_2+\beta_\ell^2+3) P_2 - (m_\ell/\sqrt{q^2}) (6\beta_\ell^2 M_1+\beta_\ell^2 +3) S_1}\frac{d\Gamma}{dq^2}\ .
}
The generalization of $\chi$ in Eq.~(\ref{P2bar}) could be expected: in the case $A_S\to 0$, the zeroes of $A_{\rm FB}$ and $P_2$ coincide, because they are both proportional to $J_{6s}$ [see Eqs.~(\ref{AFB}) and (\ref{p2})], and that is the reason for which the combination $A_{\rm FB}/P_2$ is well defined in Eqs.~(\ref{FT}), (\ref{Aim}) and (\ref{J1s})-(\ref{J9}). However, turning on $A_S$ moves the zero in $A_{\rm FB}$ away from that of $P_2$, making $A_{\rm FB}/P_2$ singular when $J_{6s}=0$. On the other hand, the expression in Eq.~(\ref{P2bar}) is regular at every point and goes to $\chi = A_{\rm FB}/P_2$ only in the limit $A_S\to 0$.

\section{Uniangular distributions}
\label{uni}

A full angular analysis designed to extract the complete set of $q^2$-dependent observables requires a good deal of statistics, and will be possible at LHCb not before an integrated luminosity $\gtrsim 10\,{\rm fb}^{-1}$ has been collected. However, certain angular observables are available from partially integrated distributions, and experimental analyses have focused to this day on uniangular distributions, leading to the set of measured observables reviewed in Section~\ref{intro}.

Starting from the full angular distribution of Eq.~(\ref{dist}), the three uniangular distributions can be obtained:
\eqa{
\frac{d^2\Gamma}{dq^2d\phi}&=& \frac{1}{8\pi} \bigg[(6J_{1s} + 3J_{1c} - 2 J_{2s} - J_{2c} + 4 J_3 \cos{2\phi} + 4 J_9 \sin{2\phi}\bigg]\\[2mm]
\frac{d^2\Gamma}{dq^2\,d\!\cos\theta_\ell}&=& \frac{1}{8} \bigg[ 6J_{1s} + 3J_{1c} +(6 J_{2s} +3 J_{2c}) \cos{2\theta_\ell} + (6 J_{6s} + 3 J_{6c}) \cos{\theta_\ell} \bigg]\\[2mm]
\frac{d^2\Gamma}{dq^2\,d\!\cos\theta_K}&=& \frac{1}{8} \bigg[ (9 J_{1s} - 3 J_{2s}) \sin^2{\theta_K} + (9 J_{1c} - 3 J_{2c}) \cos^2{\theta_K} \bigg]
}
Substituting the expressions (\ref{J1s})-(\ref{J9}) for the $J_i$ coefficients,  the uniangular distributions in the presence of scalars can be written as functions of the observables as follows:
\eqa{
\frac{d^2\Gamma}{dq^2d\phi}&=& \frac{1}{2\pi}\left[  1 - \frac{\beta_\ell}{3} \chi P_1 \cos{2\phi} + \frac{2\beta_\ell}{3} \chi P_3 \sin{2\phi}\right] \frac{d\Gamma}{dq^2}\label{dGdphi}\\[4mm]
\hspace{-2cm} \frac{d^2\Gamma}{dq^2\,d\!\cos\theta_\ell}&=& \left[  \frac{3(M_2+1)}{2(3M_2+\beta_\ell^2+3)}  - \beta_\ell \frac{4\beta_\ell^2 M_1+M_2+\beta_\ell^2+3}{8(3 M_2+\beta_\ell^2+3)} \chi - A_{\rm FB} \cos{\theta_\ell}\right. \label{dGdtl}\\[1mm]
&&\left.  -   \bigg(\frac{3\beta_\ell^2}{2(3M_2+\beta_\ell^2+3)}  + \frac{3\beta_\ell (4\beta_\ell^2 M_1 + M_2 +\beta_\ell^2+3)}{8(3M_2+\beta_\ell^2+3)} \chi \bigg) \cos{2\theta_\ell} \right] \frac{d\Gamma}{dq^2}\nn\\[4mm]
\frac{d^2\Gamma}{dq^2\,d\!\cos\theta_K}&=& \left[  \bigg( \frac{3}{2} + \frac{6\beta_\ell^2 M_1 + \beta_\ell^2 + 3}{4\beta_\ell}\chi \bigg)\cos^2{\theta_K}
-   \frac{6\beta_\ell^2 M_1 + \beta_\ell^2 + 3}{8\beta_\ell}\chi \sin^2{\theta_K} \right] \frac{d\Gamma}{dq^2}\nn\\ \label{dGdtK}
}
where $\chi$ is given by:
\eq{
\chi = \left\{
\begin{array}{l}
\displaystyle   \frac{A_{\rm FB}}{P_2}  \quad {\rm if}\quad A_S=0\ ,\\[5mm]
\displaystyle   \frac{(3+3M_2+\beta_\ell^2)A_{\rm FB} + (m_\ell/\sqrt{q^2})\, 6\beta_\ell S_1}{(3+3M_2+\beta_\ell^2)P_2 - (m_\ell/\sqrt{q^2}) (3+6\beta_\ell^2 M_1+\beta_\ell^2) S_1}  \quad {\rm if} \quad A_S\ne 0\ .
\end{array}
\right.
}\\
Note that in the limit $m_\ell\to 0$ all scalar effects disappear. The corresponding well-known expressions for the uniangular distributions in the massless case are obtained by setting $m_\ell\to 0,\ \beta_\ell\to 1$ and $M_1,M_2\to 0$.

\section{New Physics Sensitivity of the Observables}
\label{sensi}

In this section we analyze and discuss the New Physics sensitivity of the full set of observables $O_{m_\ell\ne 0}^{\rm scalars}$. In particular, we study the impact of New Physics contributions to the Wilson Coefficients:
\eq{
\C{i} = \C{i}^{SM} + \delta\C{i}\ ,
}
where $i=7^{(\prime)},9^{(\prime)},10^{(\prime)},S,P$, always taking into account the existing  bounds from other processes that constrain substantially the New Physics contributions $\delta\C{i}$.

We consider the 10 FFI observables $P_{1,2,3,4,5,6}$, $M_{1,2}$ and $S_{1,2}$ in terms of the transversity amplitudes $A_\|^{L,R}$, $A_\bot^{L,R}$, $A_t$ and $A_S$. These amplitudes can be written in terms of the  Wilson coefficients $\C{i}$ and a set of seven form factors $V(q^2)$, $A_{0,1,2}(q^2)$ and $T_{1,2,3}(q^2)$ (see for example Ref.~\cite{0008255}). We first consider the SM contribution to the observables including NLO corrections, hadronic uncertainties and an estimate of the $\Lambda/m_b$ corrections. We will see that indeed these FFI observables show reduced hadronic uncertainties.

After having the SM contribution under control, we consider NP contributions in several different scenarios, all of them compatible with current bounds from other processes, and study the possible deviations from the SM. The outcome of this analysis is shown in Figs.~\ref{fig:Ps},\ref{fig:M-S} and Tables \ref{TableZeroes},\ref{Sensitivity}.

\subsection{SM contribution and hadronic uncertainties}
\label{SMhad}

The SM Wilson coefficients at the matching scale $\mu_0=2 M_W$, and their running from $\mu_0$ down to $\mu_b=4.8\,{\rm GeV}$, as well as the running of quark masses and couplings, are obtained following Refs.~\cite{0512066,0306079,0411071,0312090,0609241}  (see also Ref.~\cite{DescotesGenon:2011yn}). For reference we quote in Table~\ref{WCSM} the used values for the Wilson coefficients at the hadronic scale, taken from Ref.~\cite{DescotesGenon:2011yn}.

\begin{table}
\begin{center}
\small
\begin{tabular}{||c|c|c|c|c|c|c|c|c|c||}
\hline\hline
$\!\C1(\mu_b)\!$ &   $\!\C2(\mu_b)\!$ &  $\!\C3(\mu_b)\!$ &  $\!\C4(\mu_b)\!$
& $\!\C5(\mu_b)\!$ & $\!\C6(\mu_b)\!$ & $\!\C7^{\rm eff}(\mu_b)\!$ & $\!\C8^{\rm eff}(\mu_b)\!$
& $\!\C9(\mu_b)\!$ & $\! \C{10}(\mu_b)\!$ \\
\hline
-0.2632 & 1.0111 & -0.0055 & -0.0806 & 0.0004 &
0.0009 &  -0.2923 & -0.1663 & 4.0749 & -4.3085\\
\hline\hline
\end{tabular}
\caption{NNLO Wilson coefficients in the Standard Model at the scale $\mu_b=4.8\,{\rm GeV}$ \cite{DescotesGenon:2011yn}. In the computation of the observables, we consider a variation of $\mu\in [\mu_b/2,2\mu_b]$. The coefficients $\C{9}$ and $\Ceff{9}$ are related though $\Ceff{9}=\C{9}+Y(q^2)$ (see Ref.~\cite{0106067}).}
\label{WCSM}
\end{center}
\end{table}

Concerning the seven $B\to K^*$ form factors ($V(q^2)$, $A_{0,1,2}(q^2)$ and $T_{1,2,3}(q^2)$), their $q^2$-dependence is parametrized following Ref.~\cite{1006.4945}, giving more conservative uncertainties than other parameterizations. Their values at $q^2=0$ are given in the same reference, obtained from light-cone sum rules with B distribution amplitudes. The definitions of the soft form factors $\xi_{\|,\bot}$ in terms of the full form factors are given in Ref.~\cite{0412400}. All numerical inputs used in this analysis are the same as the ones tabulated in Section 2.1 of Ref.~\cite{DescotesGenon:2011yn}. Maximum and minimum values for $\xi_\|(q^2)$ and $\xi_\bot(q^2)$ give rise to the grey regions in Figs.~\ref{fig:Ps},\ref{fig:M-S} around the central SM value.

At this point, the rest of the hadronic uncertainties are calculated:
\begin{enumerate}
\item The renormalization scale $\mu_b$ is varied between $2.4$ and $9.6\,{\rm GeV}$.
\item The value of $\hat m_c=m_c/m_b$ is varied in the range $\hat m_c=0.29{\pm 0.02}$, according to the discussion in Refs.~\cite{0106067,0103087}.
\item The uncertainty related to the factor that determines the relative size of the hard scattering term vs. the form factor, (see Eq.~(55) of Ref.~\cite{0106067} and discussion below), is estimated at the level of a 30\%.
\end{enumerate}
These uncertainties are added in quadrature together with the uncertainties related to the form factors, giving rise to the orange bands in Figs.~\ref{fig:Ps},\ref{fig:M-S}, on top of the gray bands (which include only the form factor uncertainties).

As a third step, $\Lambda/m_b$ contributions are estimated following the procedure in Section 2.3 of Ref.~\cite{matias2}, but widening the error band to include a $68.2\%$ of the probability (as opposed to the $66\%$ used in that reference). This uncertainty is added in quadrature to the rest of the uncertainties computed before, giving rise to the light green bands (5\% $\Lambda/m_b$ correction) and the dark green bands (10\% $\Lambda/m_b$ correction) in Figs.~\ref{fig:Ps},\ref{fig:M-S}.

\subsection{New Physics}

\begin{figure}
\begin{center}
\includegraphics[width=7.2cm,height=5.3cm]{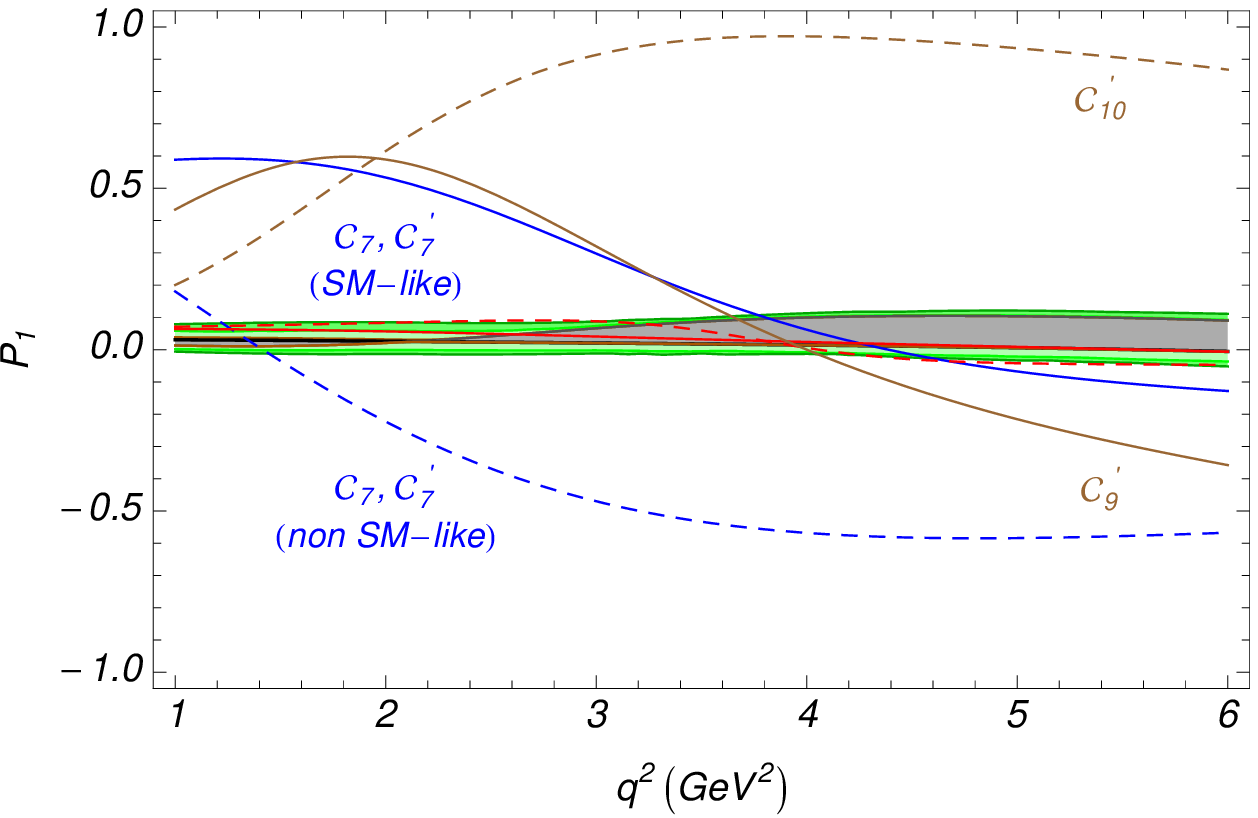}\hspace{0.7cm}
\includegraphics[width=7.2cm,height=5.3cm]{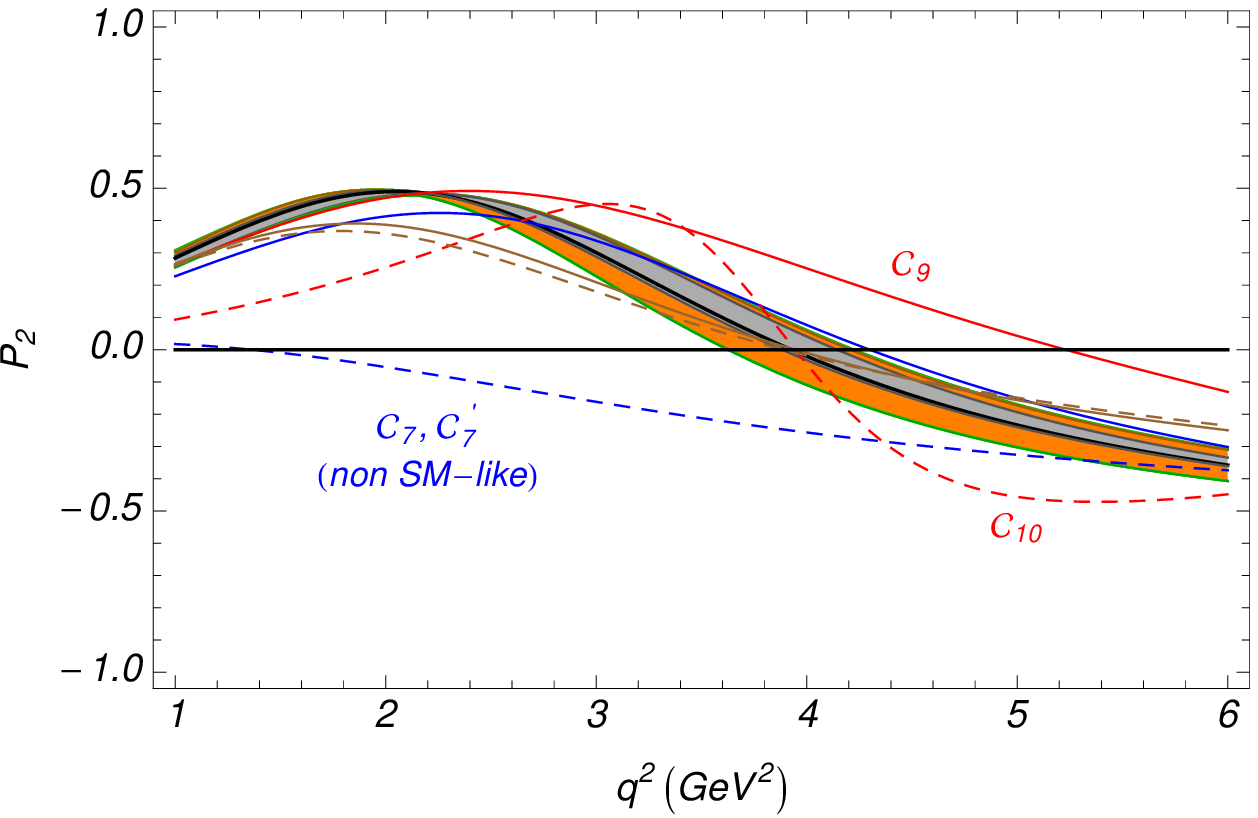}\\[4mm]
\includegraphics[width=7.2cm,height=5.3cm]{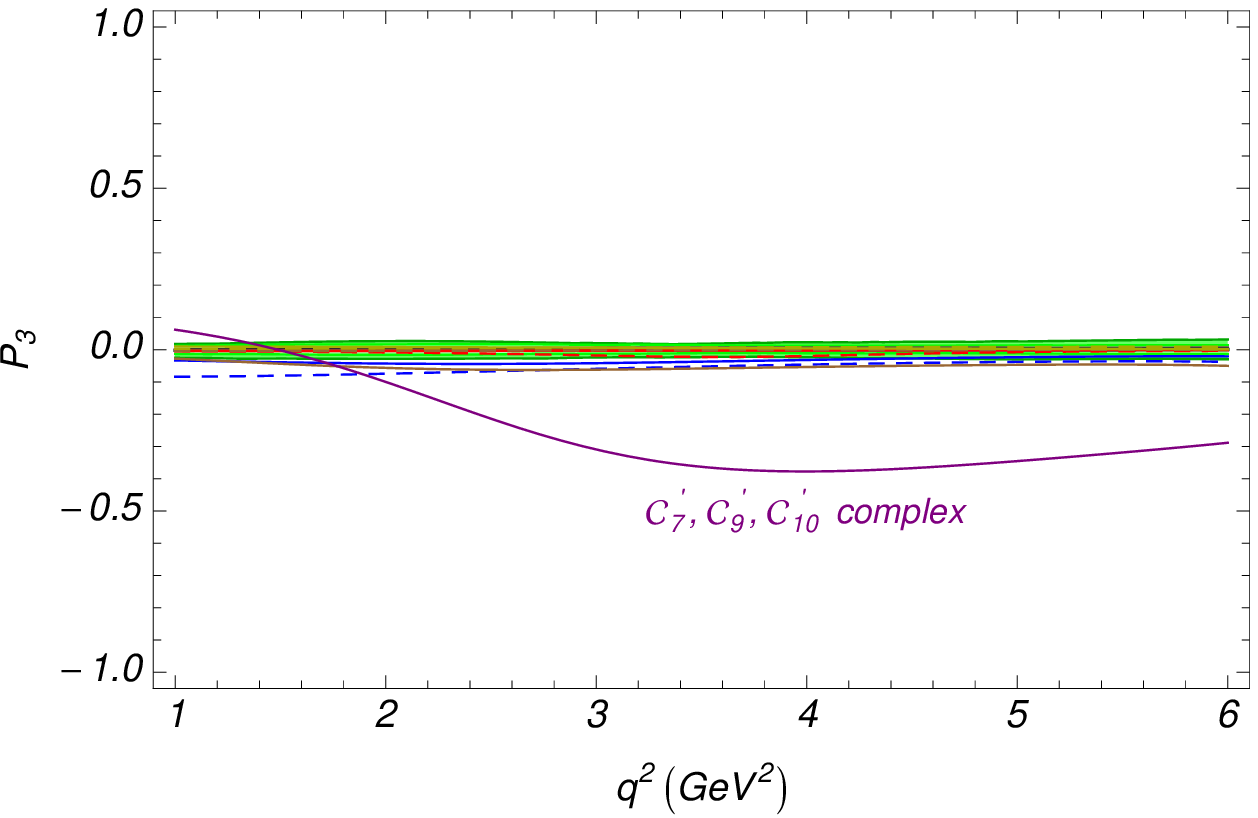}\hspace{0.7cm}
\includegraphics[width=7.2cm,height=5.3cm]{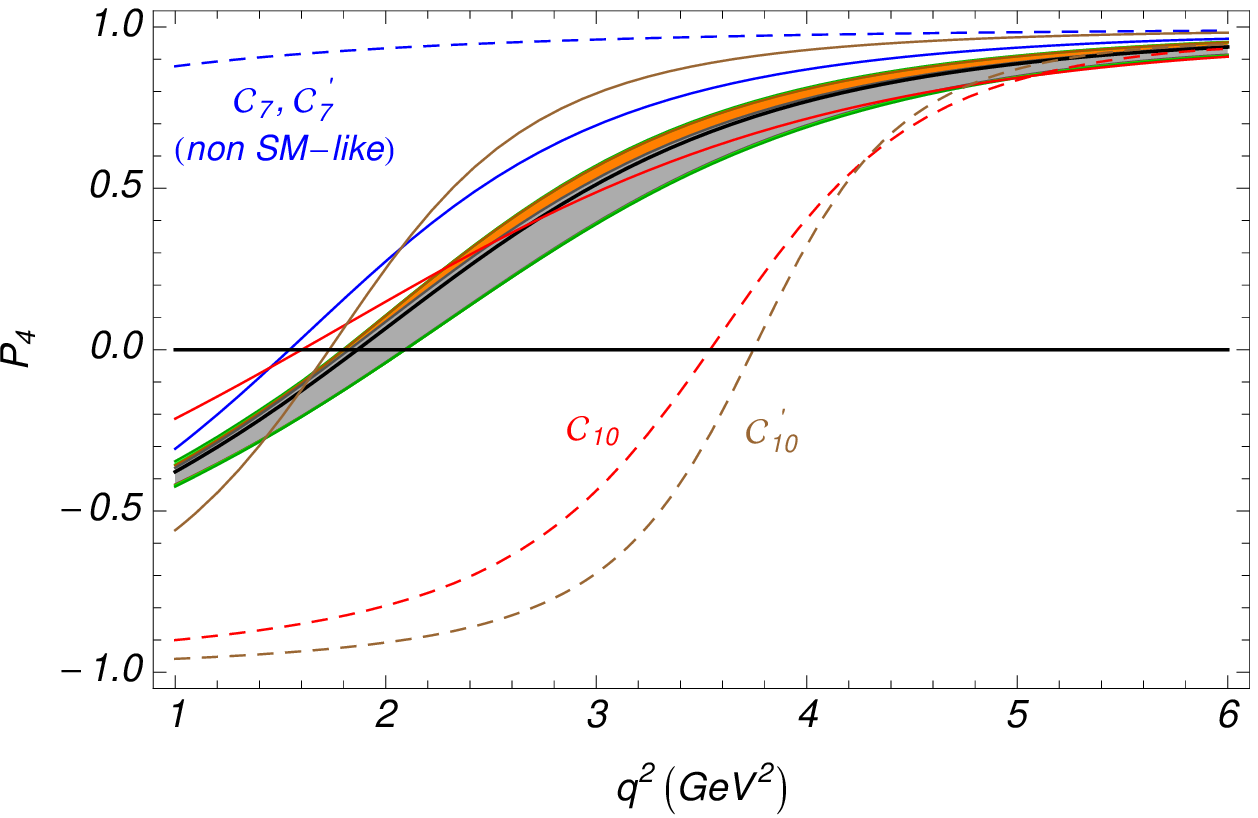}\\[4mm]
\includegraphics[width=7.2cm,height=5.3cm]{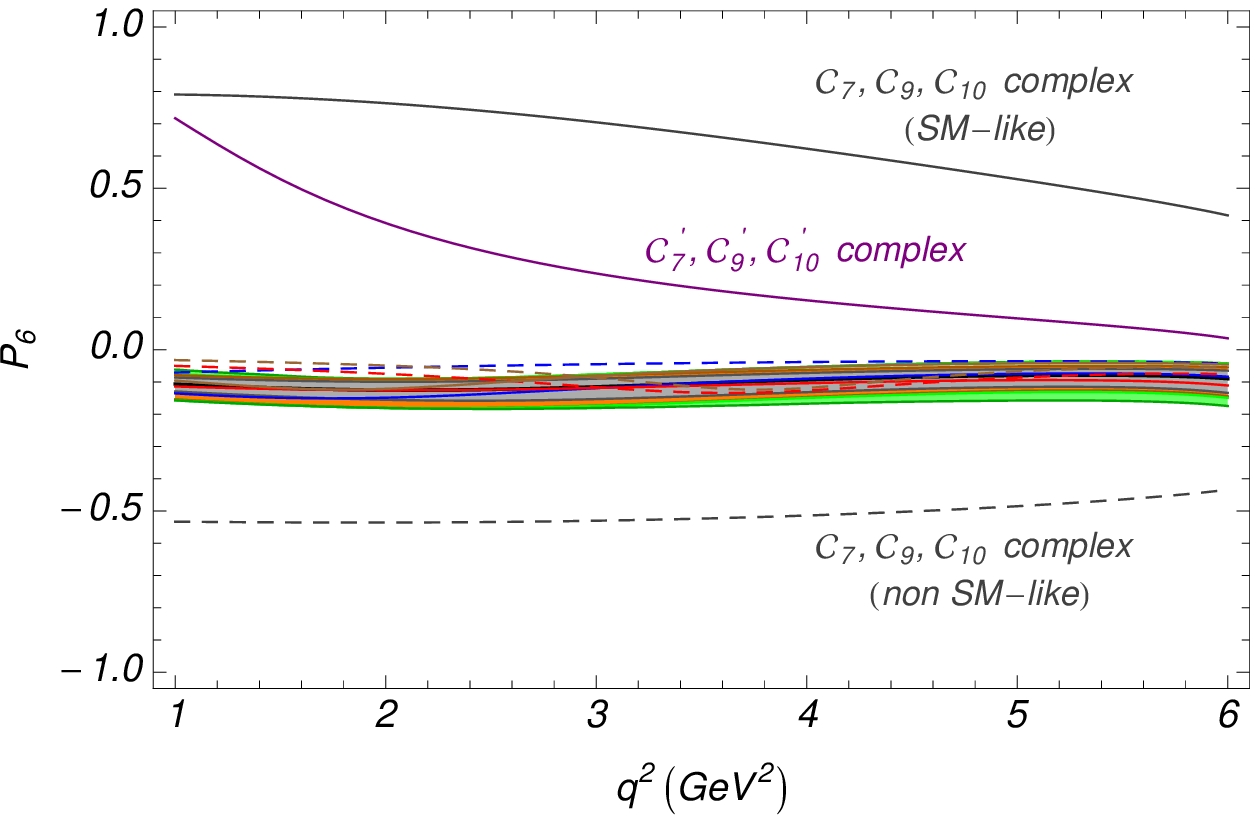}\hspace{0.7cm}
\includegraphics[width=7.2cm,height=5.3cm]{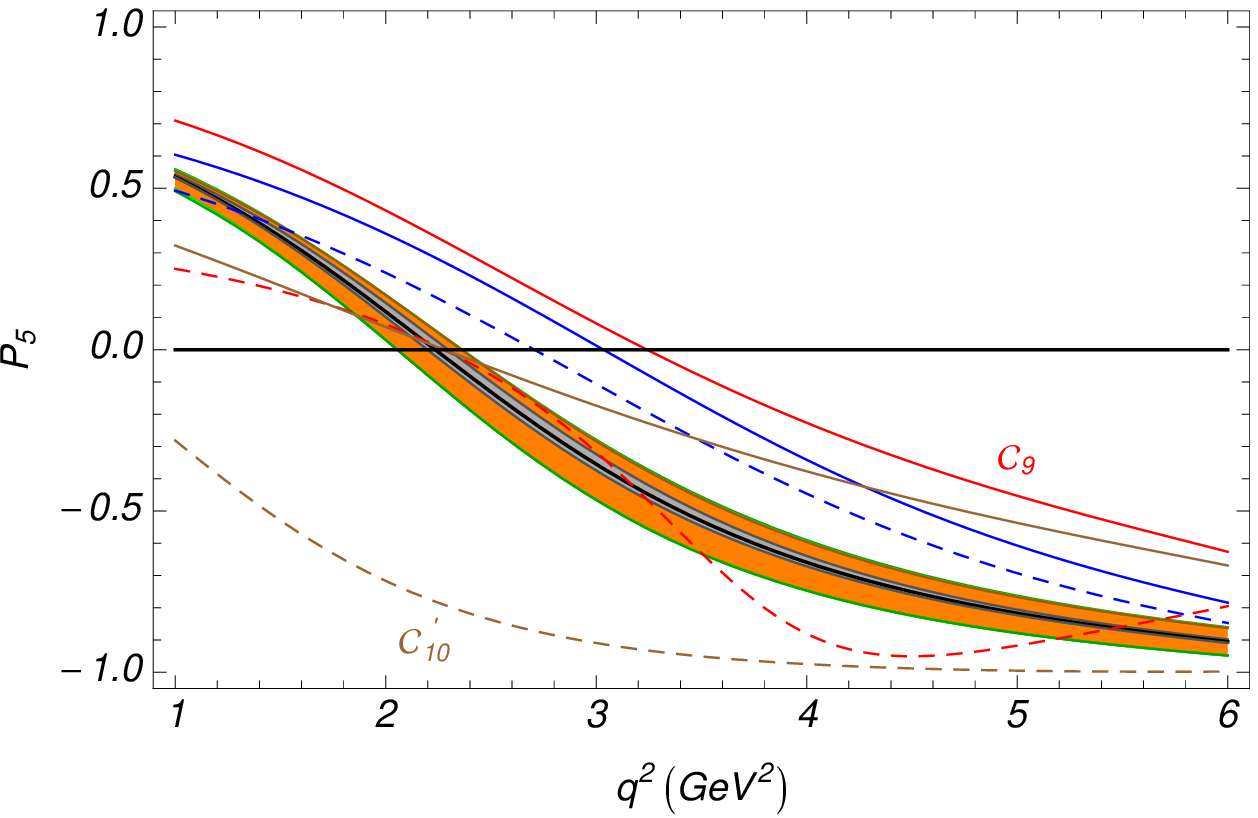}
\caption{Observables $P_{1,2,3,4,5,6}$ in the SM at NLO, including all hadronic uncertainties (wide bands) as explained in Section \ref{SMhad}. The solid and dashed curves correspond to the NP scenarios exposed in the text: Scenario A.1 (blue solid), Scenario A.2 (blue dashed), Scenario B.1 (red solid), Scenario B.2 (red dashed), Scenario C.1 (brown solid) and Scenario C.2 (brown dashed). Scenarios D.1, D.2 and D.3 are explicitly indicated in $P_3$ and $P_6$. The Wilson coefficients responsible for the largest deviations are  highlighted.}
\label{fig:Ps}
\end{center}
\end{figure}

\begin{figure}
\begin{center}
\includegraphics[width=7.2cm,height=5.3cm]{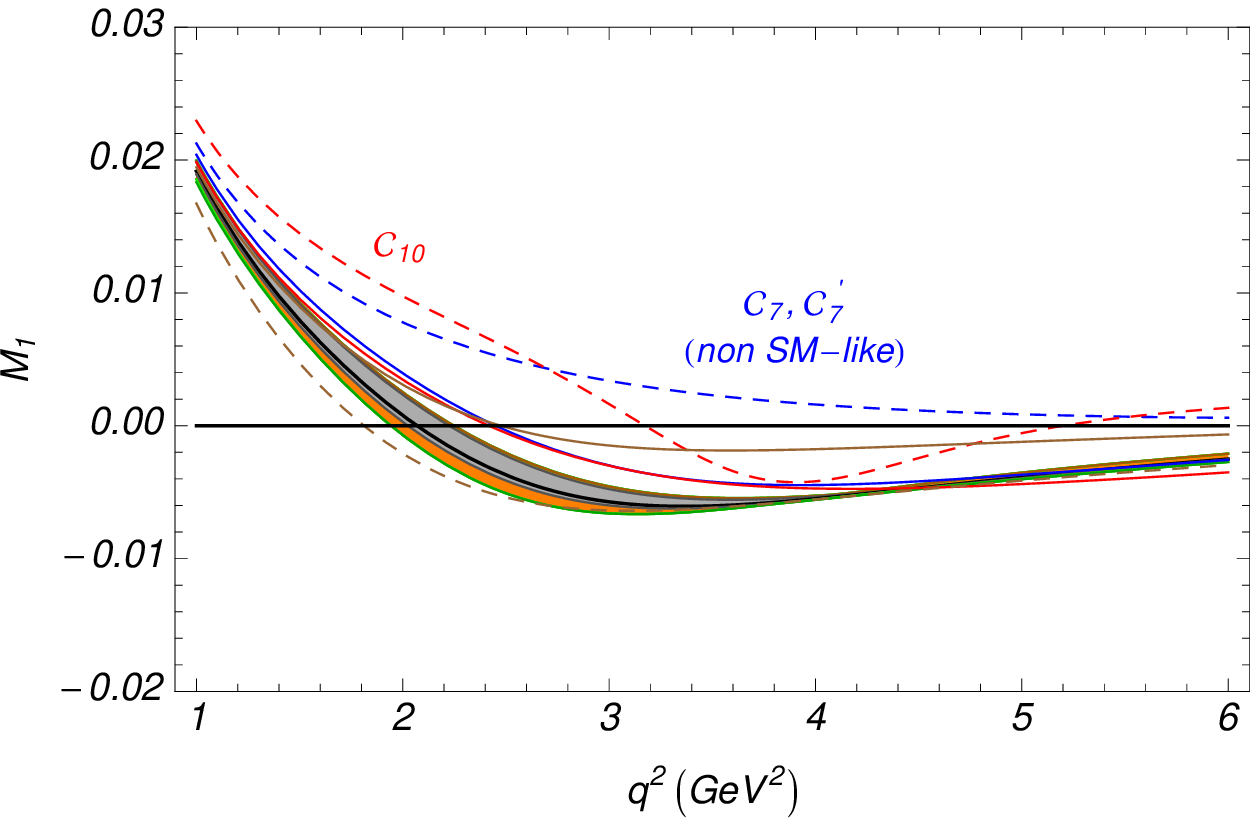}\hspace{1cm}
\includegraphics[width=7.2cm,height=5.3cm]{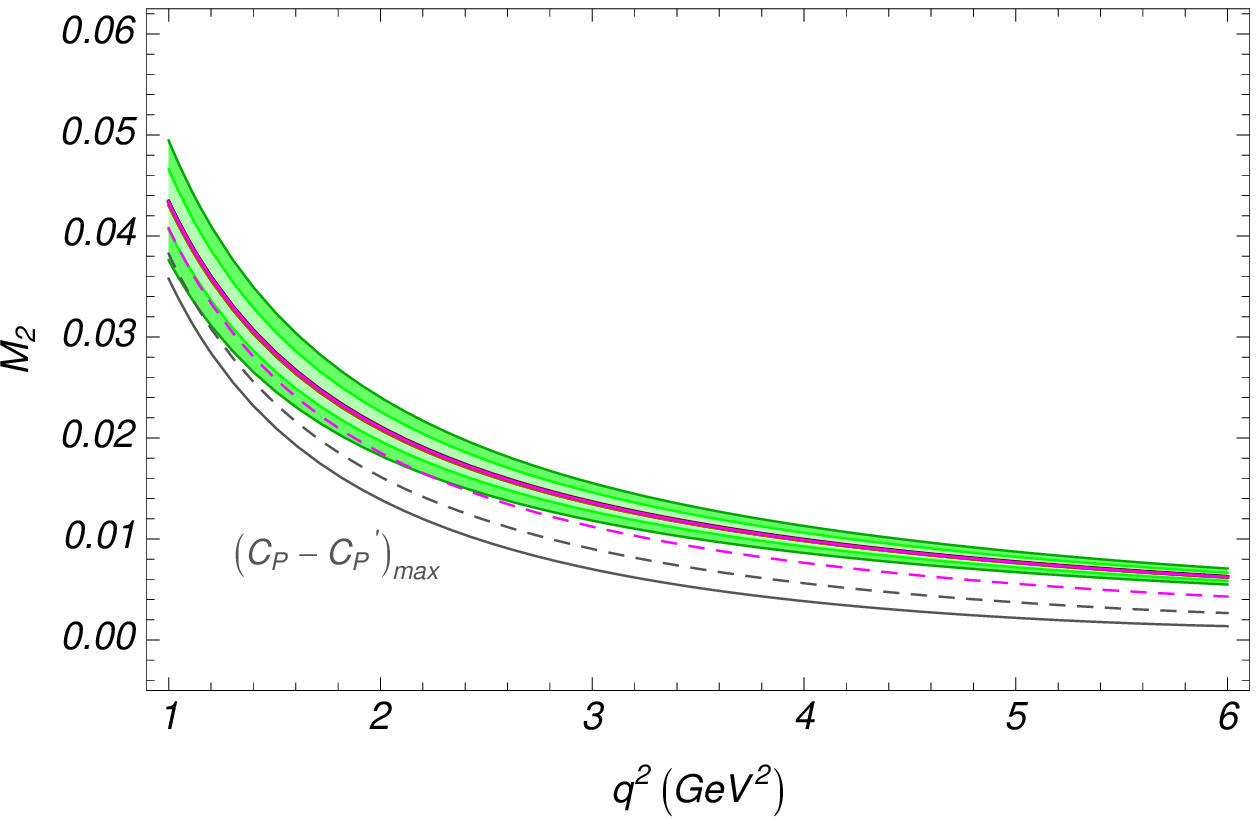}\\[4mm]
\includegraphics[width=7.2cm,height=5.3cm]{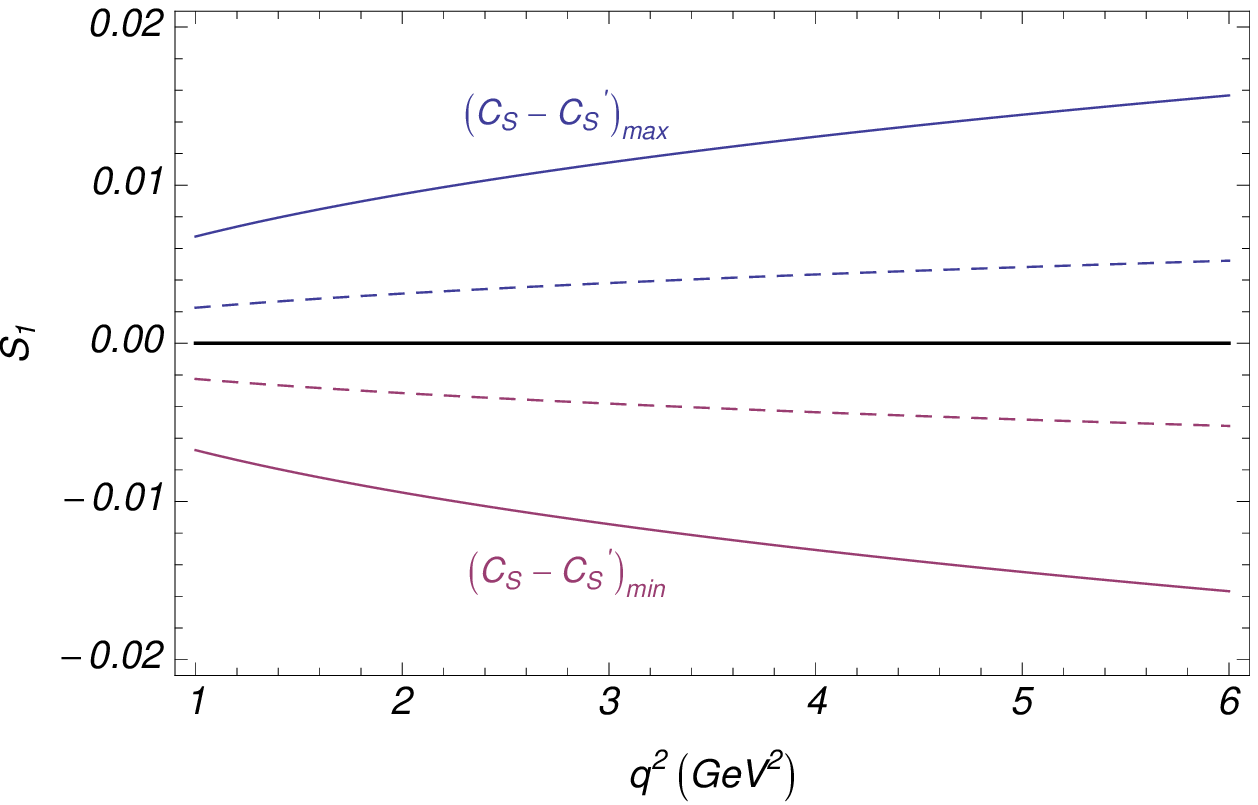}\hspace{1cm}
\includegraphics[width=7.2cm,height=5.3cm]{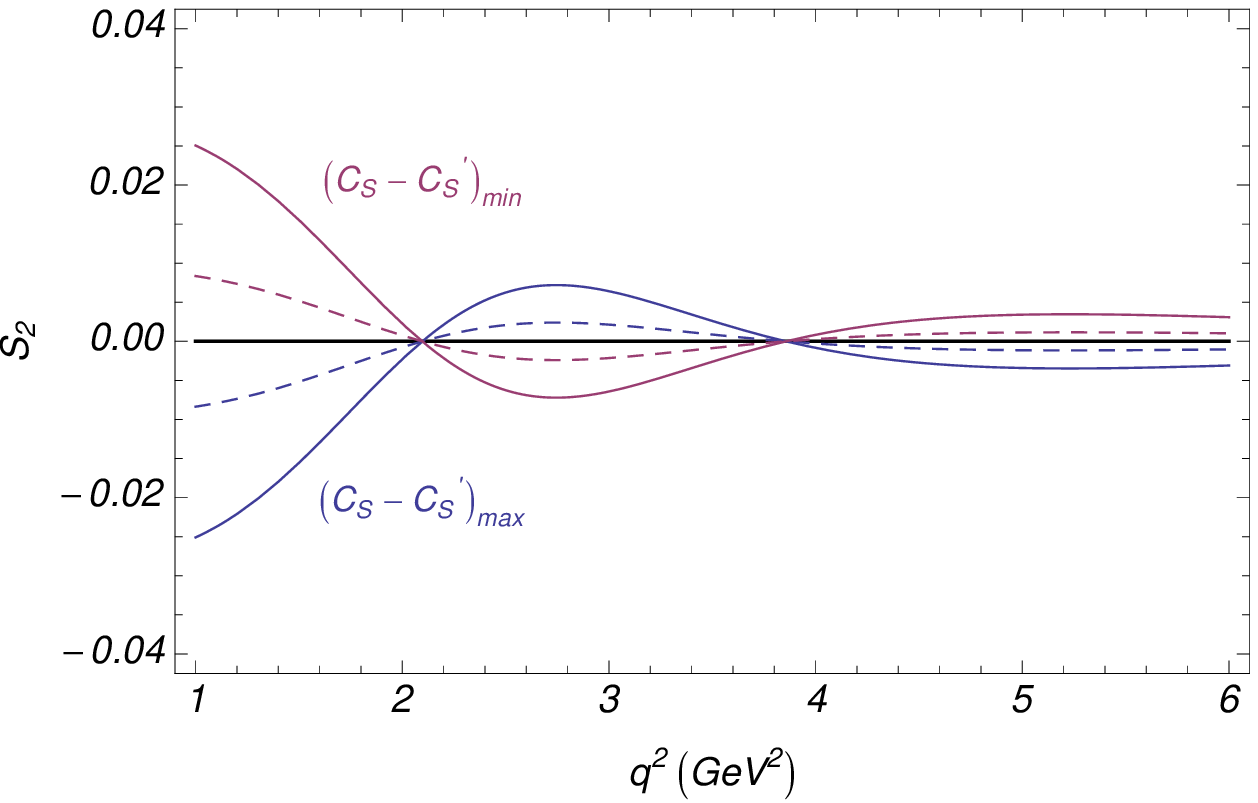}\\[4mm]
\includegraphics[width=7.2cm,height=5.3cm]{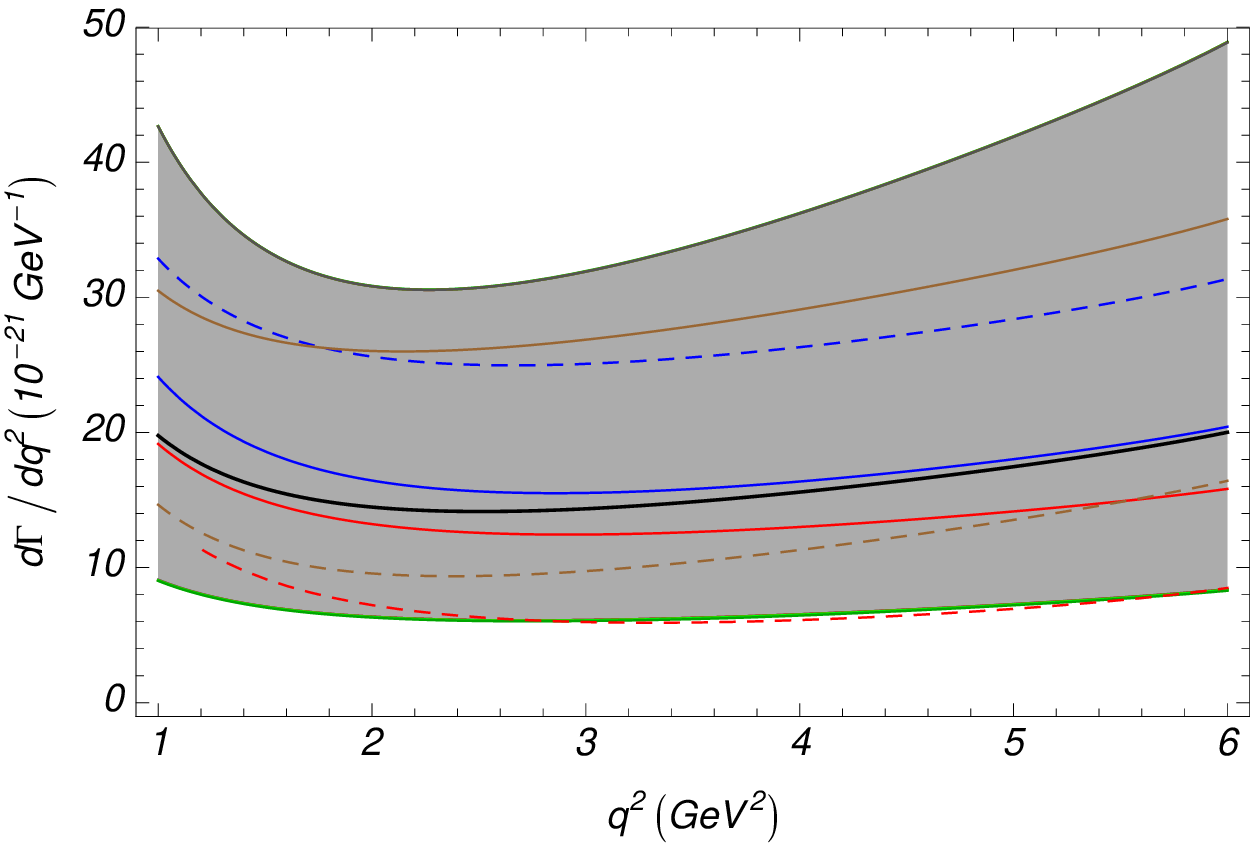}\hspace{1cm}
\includegraphics[width=7.2cm,height=5.3cm]{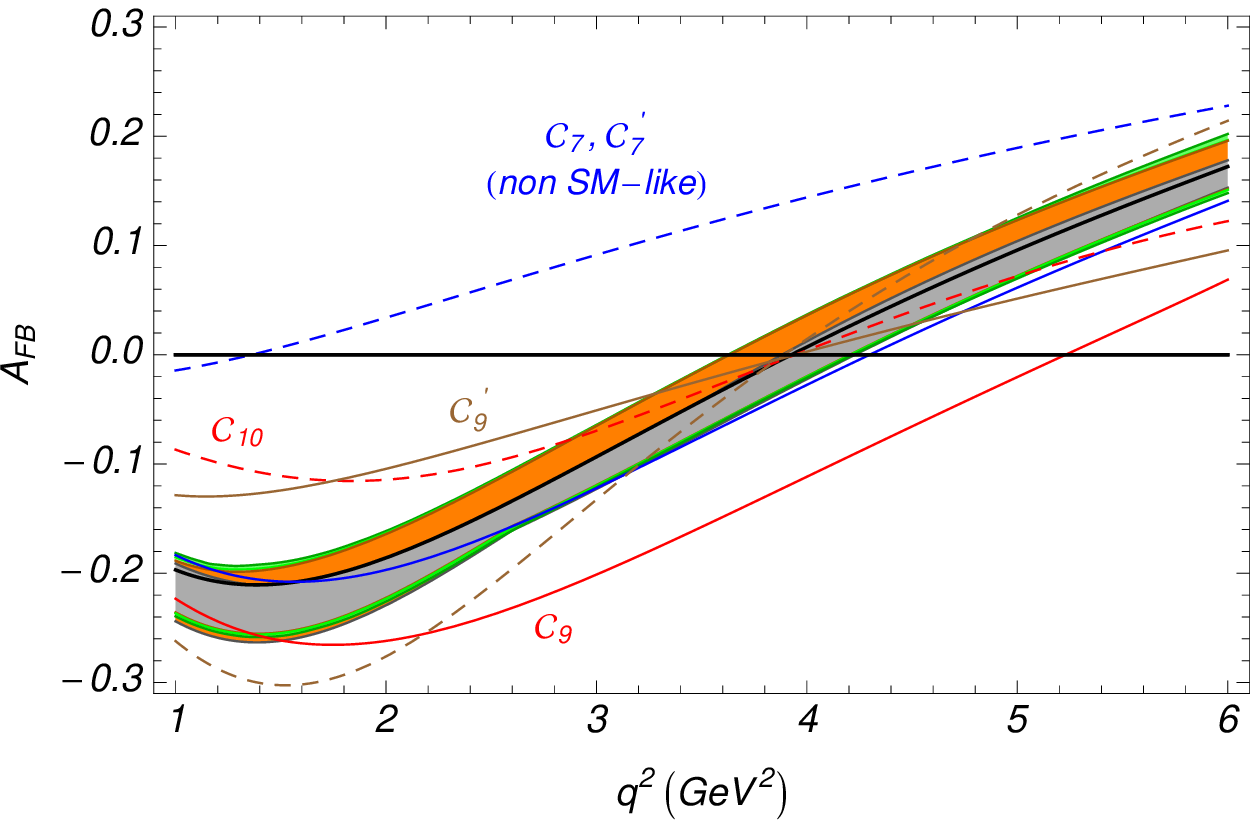}
\caption{
Observables $M_{1,2}$ (for $\ell=\mu$), $S_{1,2}$, $d\Gamma/dq^2$ and $A_{\rm FB}$. The bands correspond to the SM at NLO and with all hadronic uncertainties. Notice that $S_{1}$ and $S_2$ are strictly zero in the SM. Dashed and solid lines in $M_1$, $d\Gamma/dq^2$ and $A_{\rm FB}$ correspond to NP Scenarios A,B,C as in Fig.~\ref{fig:Ps}. In the case of $M_2$ and $S_{1,2}$, the curves correspond to: Scenario S.1 (blue solid), Scenario S.2 (blue dashed), Scenario S.3 (red dashed), Scenario S.4 (red solid), Scenario P.1 (gray solid), Scenario P.2 (gray dashed), Scenario P.3 (magenta dashed) and Scenario S.4 (magenta solid). The Wilson coefficients responsible for the largest deviations are indicated.}
\label{fig:M-S}
\end{center}
\end{figure}

The impact of NP  on $P_{1,2,3,4,5,6}$, $M_{1,2}$ and $S_{1,2}$ is shown in Figs.~\ref{fig:Ps},\ref{fig:M-S}. According to the model-independent fit of Ref.~\cite{DescotesGenon:2011yn} (updated in Ref.~\cite{Proc}), three sets of values for $\C{i}$ are chosen in order to represent the NP impact on the observables.\\[3mm]
\n {\bf Scenario A ($\C{7}$,$\,\Cp{7}$):} In this scenario, $\C{7}$ and $\Cp{7}$ are chosen according to the allowed regions obtained in the analysis of Refs.~\cite{DescotesGenon:2011yn,Proc}, while the rest of the coefficients are set to their SM values. In particular, $\Cp{7}$ is set to values where deviations between experimental data and the SM are maximal. We choose two subscenarios: Scenario A.1 corresponds to a ``SM-like'' point in the $\C{7} - \Cp{7}$ plane belonging to the allowed connected region where the SM lives (see Fig.~2 of Ref.~\cite{Proc}). Scenario A.2 corresponds to a ``non SM-like'' point belonging to a disconnected region, most likely to be probed in the near future by improved measurements of the branching ratio of $B\to X_s\mu^+\mu^-$.  The values of the relevant Wilson coefficients in these scenarios are summarized in the first column of Table~\ref{TableWCsABC}. \\[3mm]
\n {\bf Scenario B ($\C{9}$,$\,\C{10}$):} In this scenario, $\C{7}$ and $\Cp{7}$ are fixed with small NP contributions, while $\C{9}$ and $\C{10}$ take maximum allowed values compatible with the chosen $\C{7}$ and $\Cp{7}$. We distinguish between Scenario B.1, where only $\C{9}$ receives a non-zero NP contribution, and Scenario B.2 where the NP enters only in $\C{10}$. The values of the relevant Wilson coefficients in these scenarios are summarized in the second column of Table~\ref{TableWCsABC}.\\[3mm]
\n {\bf Scenario C ($\Cp{9}$,$\,\Cp{10}$):} In this scenario, $\C{7}$, $\Cp{7}$ are fixed as in Scenario B, and $\C{9}$, $\C{10}$ are SM, whereas  $\Cp{9}$ and $\Cp{10}$ take the maximum allowed values compatible with the given $\C{7}$, $\Cp{7}$, according to Refs.~\cite{DescotesGenon:2011yn,Proc}. We again distinguish between Scenario C.1, where only $\Cp{9}$ receives a non-zero NP contribution, and Scenario C.2 where the NP affects only to $\Cp{10}$. The values of the relevant Wilson coefficients in these scenarios are summarized in the third column of Table~\ref{TableWCsABC}.\\

\begin{table}[t]
\begin{center}
\begin{tabular}{||c||c|c||c|c||c|c||}
\multicolumn{1}{}{}  & \multicolumn{2}{c}{Scenario A} & \multicolumn{2}{c}{Scenario B} & \multicolumn{2}{c}{Scenario C} \\
\hline\hline
 & A.1 & A.2 & B.1 & B.2 & C.1 &C.2 \\ 
\hline\hline
$\delta\C{7}(\mu_b)$  & $-0.041$ & $0.25$ & $-0.002$ & $-0.002$ & $-0.002$ & $-0.002$\\
\hline
$\delta\Cp{7}(\mu_b)$ & $-0.114$ & $-0.414$ & $-0.006$ & $-0.006$ & $-0.006$ & $-0.006$ \\
\hline
$\delta\C{9}(\mu_b)$  & -- & -- & $-1.25$ & -- & -- & --\\
\hline
$\delta\Cp{9}(\mu_b)$ & -- & -- & -- & -- & $-3$ & -- \\
\hline
$\delta\C{10}(\mu_b)$  & -- & -- & -- & $3$ & -- & --  \\
\hline
$\delta\Cp{10}(\mu_b)$ & -- & -- & -- & -- & -- & $-3.5$ \\
\hline\hline 
\end{tabular}
\caption{Wilson Coefficients at the hadronic scale $\mu_b=4.8\,{\rm GeV}$ within Scenarios A,B,C.}
\label{TableWCsABC}
\end{center}
\end{table}

Observables $P_3$ and $P_6$ are mostly sensitive to imaginary components of the Wilson coefficients, since they are built out of imaginary parts of amplitude products [see Eqs.~(\ref{p3}) and (\ref{p6})]. In order to test this dependence, we consider a fourth scenario with complex NP contributions to the Wilson coefficients $\C{7,9,10}$ and $\Cp{7,9,10}$:\\[3mm]
\n {\bf Scenario D (complex WC's):} In this scenario, the NP contributions $\delta\C{7,9,10}^{(\prime)}$ take complex values. We consider three possibilities. In Scenarios D.1 and D.2, only $\C{7,9,10}$ receive NP contributions. Scenario D.1 consists on a point inside the ``SM-like" allowed region found in Ref.~\cite{Altmannshofer:2011gn}, while Scenario D.2 is a point in the other ``non SM-like" region of Ref.~\cite{Altmannshofer:2011gn}. In Scenario D.3, only $\Cp{7,9,10}$ are affected by  NP. The values chosen for the Wilson coefficients in these scenarios are summarized in Table~\ref{TableWCsD}.\\

\begin{table}
\begin{center}
\begin{tabular}{||c||c|c|c||}
\hline\hline
& Scenario D.1 & Scenario D.2  & Scenario D.3 \\ 
\hline\hline
$\delta\C{7}(\mu_b)$  & $0.1 + 0.5\,i$ & $1.5 + 0.3\,i$  &  -- \\
\hline
$\delta\C{9}(\mu_b)$ & $-1.4$ & $-8 + 2\,i$ & -- \\
\hline
$\delta\C{10}(\mu_b)$  & $1 - 1.5\,i$ & $8 - 2\,i$ & --\\
\hline
$\delta\Cp{7}(\mu_b)$ & -- & --  & $-0.3 - 0.1\,i$\\
\hline
$\delta\Cp{9}(\mu_b)$  & -- & --  & $3 + i$ \\
\hline
$\delta\Cp{10}(\mu_b)$ & -- & --  & $-0.6 + 2i$ \\
\hline\hline 
\end{tabular}
\caption{Wilson Coefficients at the hadronic scale $\mu_b=4.8\,{\rm GeV}$ within Scenario D.}
\label{TableWCsD}
\end{center}
\end{table}

Finally, we consider two additional scenarios with scalar and pseudoscalar New Physics contributions, to study the scalar observables $S_{1,2}$ and the pseudoscalar sensitivity of $M_2$:\\[3mm]
\n {\bf Scenario S ($\C{S}-\Cp{S}$):} All Wilson coefficients are SM except for $\C{S}^{(\prime)}$. Since the amplitudes are only sensitive to the difference $\C{S}-\Cp{S}$, we consider four different values for this difference, all compatible with the latest $B_s\to\mu^+\mu^-$ bounds \cite{Bmumu}: 
\eq{BR(B_s\to\mu^+\mu^-)< 4.5 \times 10^{-9}}
at 95\% confidence level. 
These four values constitute Scenarios S.1 to S.4 and are summarized in Table~\ref{TableWCsSP}.\\[3mm]
\n {\bf Scenario P ($\C{P}-\Cp{P}$):} In this case, all Wilson coefficients are SM except for $\C{P}^{(\prime)}$. Again, since the amplitudes are only sensitive to the difference $\C{P}-\Cp{P}$, we consider four different values compatible with the $B_s\to\mu^+\mu^-$ bound. These four values constitute Scenarios P.1 to P.4 and are summarized in Table~\ref{TableWCsSP}.\\

\begin{table}
\begin{center}
\begin{tabular}{||c||c|c|c|c||c|c|c|c||}
\multicolumn{1}{c}{} & \multicolumn{4}{c}{Scenario S} & \multicolumn{4}{c}{Scenario P} \\
\hline\hline
& S.1 & S.2 & S.3 & S.4 & P.1 & P.2 & P.3 & P.4 \\ 
\hline\hline
$(\C{S}-\Cp{S})(\mu_b)$ & $0.03$ & $0.01$ & $-0.01$ & $-0.03$  & -- & -- & -- & -- \\
\hline
$(\C{P}-\Cp{P})(\mu_b)$ & -- & -- & -- & -- & $0.07$ & $0.0467$ & $0.0233$ & $0$ \\
\hline\hline 
\end{tabular}
\caption{Wilson Coefficients at the scale $\mu_b=4.8\,{\rm GeV}$ within Scenarios S and P.}
\label{TableWCsSP}
\end{center}
\end{table}

The set of observables can be divided in two groups:  $P_{1,2,3,4,5,6}$ and $M_1$, which are only sensitive to $\C{i}$ with $i=7^{(\prime)},\,9^{(\prime)},\,10^{(\prime)}$ constitute the first group. 
In the second group we include $M_2$ and $S_{1,2}$, which are also sensitive to $\C{S}$ and $\C{P}$. In principle, $P_5$ and $P_6$ contain also $\C{S}$ [see discussion below Eq.~(\ref{As})] but the overall sensitivity, considering the present bounds on $\C{S}$,  is negligible (for this reason we do not present the curves for $P_{5,6}$ in Scenario S). We will focus on the case $\ell=\mu$ in all considerations concerning lepton mass effects.

Within the first group we have the observables $P_1$, $P_3$ and $P_6$, that are suppressed in the SM in all the $q^2$ region, but which can take sizeable non-vanishing values in specific NP scenarios, as shown in the left column of Fig.~\ref{fig:Ps}.

The rest of observables in the first group, namely $P_2$, $P_4$, $P_5$ and $M_1$ are non-vanishing already in the SM and present a non-trivial $q^2$-dependence. They all contain a zero at a value of $q^2$ within the experimentally accessible region 1-6 GeV$^2$. At LO,  these zeroes occur at the positions specified in the left column of Table~\ref{TableZeroes} (these include NLO corrections in the Wilson coefficients). At NLO, hadronic corrections shift the zeroes by amounts that can be computed in QCD factorization (see the second column in Table~\ref{TableZeroes}). In the presence of NP, the positions of the zeroes are substantially modified (moved to lower or higher $q^2$ values) as can be seen in Figs.~\ref{fig:Ps},\ref{fig:M-S}. In the third column of Table~\ref{TableZeroes}, we summarize the position of the zeroes in the most relevant NP scenarios considered above. In some NP scenarios, the zero can even disappear from the low-$q^2$ region. This is the case for example for $P_4$ in Scenario A.2, or $P_5$ in Scenario C.2.

As mentioned before, these zeroes cannot produce any singular behavior in the coefficients $J_i$. The only potential singular points are the zero of $P_2$ and the zero of the combination $3+\beta_\ell + 3M_2$. However, $3+\beta_\ell + 3M_2$ is always strictly positive, and the zero of $P_2$ coincides exactly with the zero of the forward-backward asymmetry, and cancels out. This is true in the absence of scalars; in general, the relevant parameter is $\chi$, which is always well defined, as discussed in Section~\ref{sec:scalar}.

\begin{table}
\begin{center}
\begin{tabular}{||c||r|c||c||c||}
\hline\hline
Obs. & \multicolumn{2}{|c|}{$q_0^{2\, SM}$ at Large Recoil} & $q_0^{2\, SM}$ (NLO) & $q_0^{2\, NP}$ \\
\hline\hline
$P_2$ & \begin{minipage}{2.4cm} \vspace{1.5mm} $-\ds \frac{2m_b m_B\,\Ceff{7}}{\Ceff{9}}$ \vspace{0.5mm} \end{minipage} & \hspace{0.3cm} 3.06\hspace{0.3cm}  & 3.93  &  $5.23$ [B.1] \\
\hline
$P_4$ & \begin{minipage}{5.9cm} \vspace{1.5mm} $-\ds \frac{2m_b m_B(m_B\,\Ceff{9} + 2m_b\,\Ceff{7})\,\Ceff{7}}{2 m_b\,\Ceff{7}\,\Ceff{9} + m_B\,({\Ceff{9}}^2+\C{10}^2)}$ \vspace{2mm} \end{minipage} & 1.58 & 1.87  & \begin{minipage}{1.8cm}
$3.75$ [C.2]\\
$3.55$ [B.2]
\end{minipage}\\
\hline
$P_5$ & \begin{minipage}{3.5cm} \vspace{1.5mm} $-\ds \frac{m_b m_B^2\,\Ceff{7}}{m_b\,\Ceff{7} + m_B\ \Ceff{9}}$ \vspace{2mm} \end{minipage} & 1.64 & 2.23 &
\begin{minipage}{1.8cm}
\vspace{2mm}
$3.25$ [B.1]\\
$3.03$ [A.1]\\[-3mm]
\end{minipage}\\
\hline
$M_1$ & \begin{minipage}{2.5cm} \vspace{1.5mm} $-\ds \frac{2m_b m_B\,\Ceff{7}}{\Ceff{9}-\C{10}}$ \vspace{0.5mm} \end{minipage} & 1.61 & 2.07 & $3.15$ [B.2]\\
\hline\hline 
\end{tabular}
\caption{Position of the zeroes of observables $P_{2,4,5}$ and $M_1$, in the SM at large recoil, at NLO, and in selected NP scenarios. In the last column, the zeroes correspond to the scenarios indicated in brackets. All values are given in GeV$^{\,2}$. In the calculation of the large recoil zeroes, we use the Wilson coefficients given in Table~\ref{WCSM}, and $m_b=4.68\,{\rm GeV}$. The zeroes of $A_{\rm FB}$ coincide exactly with those of $P_2$, except in the presence of scalars.}
\label{TableZeroes}
\end{center}
\end{table}

The observables $M_2$, $S_1$ and $S_2$ are affected by scalar Wilson coefficients (see Fig.~\ref{fig:M-S}) --
we recall that $\C{S}$ effects in  $P_5$ and $P_6$ are negligible--. Moreover, $M_2$ plays a special role since it is  the only observable in $\bar B\to \bar K^*\ell\ell$ sensitive to $\C{P}$. As can be seen in Fig.~\ref{fig:M-S}, the most promising place to look for scalar effects is in the observable $S_1$, when integrated in the full $q^2$ region. This is due to the fact that for fixed $\C{S}^{(\prime)}$, the curves are always positive (if $\C{S}>\Cp{S}$) or negative (if $\C{S}<\Cp{S}$), while in the absence of scalars $S_1$ is strictly zero in the full range. Integrating over $q^2$ has a clear experimental advantage from the point of view of statistics. However, we would need to discriminate values for the integrated observable below the level of $|\int dq^2 S_1| \sim 0.12\ {\rm GeV}^2$ in order to improve the $B_s\to\mu^+\mu^-$ bound.

In the case of $M_2$, considering the fully integrated  $q^2$ region also leads to better sensitivity to $\C{P}^{(\prime)}$, since $M_2>0$ for all $q^2$. However, the sensitivity in this case should be better than $\Delta(\int dq^2 M_2) \sim 0.03\ {\rm GeV}^2$, before the current bounds on the difference $\C{P}-\Cp{P}$ can be probed.

Finally, for the sake of completeness, we show in the third row of Fig.~\ref{fig:M-S} the observables $d\Gamma/dq^2$ and $A_{\rm FB}$ in the SM and in Scenarios A, B, C and D. It becomes manifest that these observables are affected by larger hadronic uncertainties, and show a milder sensitivity to  NP effects. Although the forward-backward asymmetry seems to suffer a significant enhancement in Scenario A.2, this scenario can be much more effectively probed by the observables $P_1$ or $P_4$.

\begin{table}
\begin{center}
\begin{tabular}{||c|c||}
\hline\hline
Observable & Wilson Coefficients\\
\hline\hline
$P_1$ &   $\C{7},\Cp{7},\Cp{9},\Cp{10}$  \\
\hline
$P_2$ & $\C{7},\Cp{7}$  \\
\hline
$P_3$ &  $\im(\Cp{7},\Cp{9},\Cp{10})$  \\
\hline
$P_4$ & $\C{7},\Cp{7},\C{10},\Cp{10}$  \\
\hline
$P_5$ &  $\Cp{10}$, [$\C{9}$]\\
\hline
$P_6$ &  $\im(\C{7}^{(\prime)},\C{9}^{(\prime)},\C{10}^{(\prime)})$  \\
\hline
$M_1$ & [$\C{7},\Cp{7}$] \\
\hline
$M_2$ &  [$\C{P}-\Cp{P}$]  \\
\hline
$S_1$ &  $\C{S}-\Cp{S}$ \\
\hline
$S_2$ &  [$\C{S}-\Cp{S}$]  \\
\hline\hline 
\end{tabular}
\caption{Main contributions to the observables from NP Wilson Coefficients. The listed WC's produce strong deviations from the SM (consistent with all other bounds). For those listed in brackets, the effect is milder.}
\label{Sensitivity}
\end{center}
\end{table}

A summary of the NP sensitivity of each observable can be found in Table \ref{Sensitivity}. For each observable, we list the Wilson coefficients whose NP contributions affect  most strongly the values of the given observable. We also present within brackets those Wilson coefficients whose effect is moderate.

\section{Summary of results}
\label{sec:sum}

The angular distribution of the four body decay $\bar B_d\to \bar K^{*0}(\to K \pi) \ell^+\ell^-$ can be studied experimentally by doing a fit to the coefficients $J_i(q^2)$ of the distribution, defined customarily as in Eq.~(\ref{dist}). The contact with theory is given by the expressions of the coefficients $J_i$ in terms of the transversity amplitudes, as shown in Eq.~(\ref{Js}) [where $\beta_\ell$ is defined in Eq.~(\ref{beta})]. In general, these amplitudes are $A^{L,R}_{\|,\bot,0}$, $A_t$ and $A_S$.

However, depending on the case (if the masses of the leptons are negligible, as for example if $\ell^\pm=e^\pm$, or if there are no NP contributions from scalar operators), these coefficients are not independent. In such cases, an independent fit to all the coefficients can be problematic. Moreover, since such correlations contain physical information, it is interesting not only to have them identified, but to take profit of them.

On the other hand, the coefficients $J_i$ are not the best observables to consider from the theory point of view because they suffer from large hadronic uncertainties. This has been noticed before and many theoretically clean observables have been devised in the literature. However, not all the observables that one can construct from the transversity amplitudes can be extracted from the angular distribution if they violate some ``symmetry properties''.

With the development of a formalism based on these ``symmetries'', together with the above considerations, in this paper we have constructed a complete and efficient set of observables engineered to extract the maximum information from the angular distribution:

\begin{enumerate}

\item In the most general case, the chosen basis of observables is composed by the FFD observables $d\Gamma/dq^2$ and $A_{\rm FB}$ [Eqs.~(\ref{dgamma}),(\ref{AFBsc})], and the FFI observables $P_{1,2,3,4,5,6}$ [Eqs.~(\ref{p1})-(\ref{p6})], $M_{1,2}$ [Eqs.~(\ref{m1}),(\ref{m2})] and $S_{1,2}$ [Eqs.~(\ref{s1}),(\ref{s2})]. The angular distribution in terms of the observables is given by Eqs.~(\ref{J1s})-(\ref{J9}), with $\chi$ defined in Eq.~(\ref{P2bar}). The uniangular distributions can be found in Eqs.~(\ref{dGdphi})-(\ref{dGdtK}).

\item The reduction to the $A_S=0$ case is obtained by setting $S_i=0$ and $\chi=A_{\rm FB}/P_2$. The vanishing of $S_i$ leads to two dependencies among the $J$'s: $J_{6c}= 0$ and the relationship of Eq.~(\ref{relation}). In fact, $S_1$ and $S_2$ measure the breaking of these relations by scalar effects.

\item The reduction to the massless case ($m_\ell=0$) is obtained by setting $\beta_\ell\to 1$ and $M_i=0$. This leads to two further relationships between the $J$'s: $J_{1s}=3J_{2s}$ and $J_{1c}=-J_{2c}$. In fact, $M_1$ and $M_2$ measure the breaking of these relations by mass effects.

\end{enumerate}

The NP sensitivity analysis shows that these observables are quite sensitive to complementary NP effects. This can be observed in Figs.~\ref{fig:Ps},\ref{fig:M-S} and in Table~\ref{Sensitivity}. It is almost a certainty that future analyses of LHC data by the LHCb collaboration will be putting serious constraints on NP physics by studying these observables, or else discrepancies with respect to our SM predictions will be made manifest and constitute part of the first studies of true physics beyond the Standard Model.

\subsubsection*{Acknowledgements}

We would like to thank Damir Becirevic for correspondence, as well as Tobias Hurth and Nazila Mahmoudi for pointing out a missing factor of 2 in Eqs.~(\ref{At}) and (\ref{As}). J.M. acknowledges financial support from FPA2011-25948, SGR2009-00894.
F.M. acknowledges financial support from FPA2010-20807 and the Consolider CPAN project. J.V. is supported in part by ICREA-Academia funds and FPA2011-25948.

\newpage

\appendix

\section{Symmetry Formalism}
\label{appA}

In this appendix we complete the symmetry approach to the angular distribution that was originally presented for the massless case in Ref.~\cite{matias2}. We present two different formalisms to describe the distribution. The first formalism, constructed using unitary matrices and two-component complex vectors, will be appropriate to describe both the massless and massive cases. However, in order to introduce the scalar contributions a more general formalism is required. This second more powerful formalism introduces, instead,  ortogonal matrices and four-component vectors and is valid for all cases.

We follow here a bottom-up approach, from the simplest (massless) case to the general case (massive with scalars). We also recall the solution of the system, expressing transversity amplitudes in terms of $J$'s, in the massless case (see Ref.~\cite{matias2}), while the full solution to the system in the general case will be presented elsewhere \cite{jo}.

The  importance of determining these symmetries is mainly twofold. On the one hand, from the experimental point of view, the symmetries allow to identify all correlations between the coefficients of the distribution that may affect the stability of the fit; but they are also helpful  to determine which amplitudes can be consistently put to zero, in order to simplify the system and consequently the fit. On the other hand, they provide you with an alternative procedure to construct observables directly in terms of the transversity amplitudes: verifying that they are invariant in the first place, and afterwards, translating their expression in terms of transversity amplitudes to an expression in terms of the measured coefficients $J_i$ of the distribution (an example of this procedure was the observable $A_T^{(5)}$ designed in \cite{matias2}).

\subsection{Symmetries of the massless distribution}
\label{appmassless}

In this section we review the symmetry formalism for the massless angular distribution, as presented originally in Ref.~\cite{matias2}. 

The six complex amplitudes present in this case can be arranged into three complex vectors:
\eq{
n_\|=\binom{A_\|^L}{A_\|^{R*}}\ ,\quad
n_\bot=\binom{A_\bot^L}{-A_\bot^{R*}}\ ,\quad
n_0=\binom{A_0^L}{A_0^{R*}}\ .
}
All the coefficients $J_i$ can be expressed in terms of the products $n_i^\dagger\,n_j$:
\eqa{
J_{1s} & = & \frac34\left( |n_\bot|^2+ |n_\||^2\right),\hspace{0.9cm}
J_{1c}  =|n_0|^2 \,, \hspace{3cm}
J_{2s}  = \frac14 \left(|n_\bot|^2+ |n_\||^2 \right) \,, \nn\\[1mm]
J_{2c} & = & -|n_0|^2 \,,\hspace{2.85cm}
J_3  = \frac12 \left( |n_\bot|^2 - |n_\||^2 \right)\,, \hspace{0.9cm}
J_4  = \frac1{\sqrt{2}} \re (n_0^\dagger\,n_\|)\,,\nn\\[1mm]
J_5 & = &  \sqrt{2}\, \re (n_0^\dagger\,n_\bot)\,, \hspace{1.45cm}
J_{6s}  = 2\, \re (n_\bot^\dagger\,n_\|)\,, \hspace{1.85cm}
J_{7} =  -\sqrt{2}\, \im (n_0^\dagger\,n_\|)\,,\nn \\[1mm]
J_8 & = & -\frac1{\sqrt{2}} \im (n_0^\dagger\,n_\bot) \,,\hspace{1.25cm}
J_9 = -\im (n_\bot^\dagger\,n_\|)\,,\hspace{1.65cm}
J_{6c} = 0\,.
\label{Jsmassless}}
A symmetry of the angular distribution will therefore be a unitary transformation $U$ acting in the same way on $n_0$, $n_\|$ and $n_\bot$, that is: $n_i\to U n_i$. Such a symmetry has four independent parameters, and can be written as:
\eq{
  n_i^{'} = U n_i=
  \left[
    \begin{array}{ll}
      e^{i\phi_L} & 0 \\
      0 & e^{-i \phi_R}
    \end{array}
  \right]
  \left[
    \begin{array}{rr}
      \cos \theta & -\sin \theta \\
      \sin \theta &  \cos \theta
    \end{array}
  \right]
  \left[
    \begin{array}{rr}
      \cosh i \tilde{\theta} &  -\sinh i \tilde{\theta} \\
      - \sinh i \tilde{\theta} & \cosh i \tilde{\theta}
    \end{array}
  \right]
  n_i \,.
 \label{symmassless}}
Of course, other parametrizations are possible, but we prefer to keep this one to make an easy contact with the generalization to the massive case and the notation introduced in Ref.~\cite{matias2}.
%
The matrix $U$ defines the four symmetries of the massless angular distribution: two global phase transformations ($\phi_L$ and $\phi_R$), a rotation $\theta$ among the real and imaginary components of the amplitudes independently and another rotation $\tilde\theta$ that mixes real and imaginary components of the transversity amplitudes.

\subsubsection{Solution to the massless distribution}
\label{solmassless}

We can now use these symmetries to reduce the number of theoretical parameters and solve for the transversity amplitudes in terms of the coefficients $J$'s.  It is instructive to use only three out of the four symmetries and see how the extra freedom related to the fourth symmetry arises from the equations. This extra freedom gives rise to the non-linear relation between the $J$'s given in Eq.~(\ref{relation}).

Using the symmetries we choose to fix the following amplitudes to zero: $A_\|^L=0$ and $\im A_\|^R=0$. We achieve this configuration easily using the rotation phases  $\phi_L$ and $\phi_R$ to set the phases of  $A_\|^L$ and $A_\|^R$  to zero. Then a rotation by an angle $\theta$ given by 
\eq{\tan \theta= \frac{\re A_\|^L}{ \re A_\|^R}}
will also set the modulus of $A_\|^L$ to zero.

With these simplifications, rewriting the products $n_i^\dagger\,n_j$ in this configuration, and taking into account Eqs.~(\ref{Jsmassless}), one gets immediately:
\eqa{\label{aintermsJI}
A_\|^L=0 \,,\hspace{1cm} A_\|^R=\sqrt{2 J_{2s}-J_3}\ , \nn \\
A_\perp^R=-\frac{J_{6s}-2 i J_9}{2 \sqrt{2 J_{2s}-J_3}}  \,,\hspace{1cm} A_0^R=\frac{2 J_4 - i J_7}{\sqrt{4 J_{2s} - 2 J_3}}\ .
}
The remaining equations involving the last two amplitudes ($A_\perp^L$ and $A_0^L$) lead to \cite{matias2}
\begin{equation}\label{phase} e^{i (\phi_0^L  - \phi_\perp^L)}=\frac{2 (2 J_{2s}-J_3)(J_5 + 2 i J_8) - (2 J_4 + i J_7) (J_{6s}-2 i J_9)}{\sqrt{16 J_{2s}^2 - 4 J_3^2 - J_{6s}^2-4 J_9^2} \sqrt{2 J_{1c} (2 J_{2s}-J_3) - 4 J_4^2 -J_7^2}}\ ,
\end{equation}
where $\phi_0^L$ and $\phi_\perp^L$ are the phases associated to the amplitudes $A_0^L$ and $A_\perp^L$. The relation between the $J_i$ coefficients (Eq.~(\ref{relation})) arises naturally from imposing in Eq.~(\ref{phase}) that the modulus of this phase difference should be one. Notice also that the freedom to choose one of the two phases in Eq.~(\ref{phase}) to be zero  is somehow related to the freedom associated to the last unused symmetry transformation $\tilde\theta$. The choice $\phi_\perp^L=0$  fixes the last two amplitudes to
\eqa{
A_\perp^L&=&\frac{\sqrt{16 J_{2s}^2 - 4 J_3^2 - J_{6s}^2- 4 J_9^2}}{2 \sqrt{2 J_{2s}- J_3}}, \nn \\
A_0^L&=&\frac{2 (2 J_{2s}-J_3)(J_5 + 2 i J_8) - (2 J_4 + i J_7) (J_{6s} - 2 i J_9)}{\sqrt{4 J_{2s}-2 J_3}\sqrt{16 J_{2s}^2- 4 J_3^2 - J_{6s}^2 - 4 J_9^2}}.
\label{aintermsJII}}
The solution in any other configuration can be obtained by applying the symmetry transformation in Eq.~(\ref{symmassless}). Any observable constructed from the transversity amplitudes, can be expressed in terms of the coefficients $J_i$ using Eqs.~(\ref{aintermsJI}) and (\ref{aintermsJII}). The condition that the observable is invariant under the symmetries of the angular distribution, guarantees that the configuration used to derive these equations gives the same result for the observables as any other configuration, and the result is unique.

\subsection{Symmetries of the massive distribution}
\label{appmassive}

In the massive case the balance equation between theoretical and experimental degrees of freedom \cite{matias2}
\eq{n_{exp}\equiv n_J-n_d=2 n_A-n_S}
is fulfilled with $ n_J=12$, $ n_d=2$, $ n_A=7 $ and $ n_S=4 $, where $n_d$ is the number of relationships among the $J_i$, as explained in Section~\ref{sec:2}.
When the masses are switched on to account for lepton mass corrections in $B_d \to K^{*0} (\to K \pi) \ell^+ \ell^-$, all the massless symmetries described in the previous section are broken by the mass terms and have to be redefined. The four symmetries in the massive case are then:

\begin{itemize}

\item A common global phase transformation for both left and right components 
\eq{
  n_i^{'} = U_0 (\phi) n_i=
  \left[
    \begin{array}{ll}
      e^{i\phi} & 0 \\
      0 & e^{-i \phi}
    \end{array}\right] n_i \,, 
\label{symmass}}
for $i=\|,\perp,0$.
 
\item An independent new global phase transformation for the $A_t$ amplitude: $A_t^\prime= e^{i \phi_t} A_t$.

\item Two rotations $U_1(\theta)$ and $U_2(\tilde\theta)$ between the real and imaginary components  of the transversity amplitudes $A_i^{L,R}$: $$n_i^\prime=U_{1}(\theta) n_i \,, \quad \quad  \quad n_i^\prime=U_{2}(\tilde\theta) n_i \,, $$ ($i=\perp,\|,0$) with a similar structure  to those in the massless case, but including some important differences to be discussed below.

\end{itemize}

In order to find the explicit form of the last two symmetries, $U_1(\theta)$ and $U_2(\tilde \theta)$, it is helpful to analyze their infinitesimal form. This can be obtained following the approach described in Ref.~\cite{matias2}. Let us focus on $U_1(\theta)$ and require that the infinitessimal transformation of the amplitudes is a symmetry of the distribution. This leads to the following form of the infinitessimal transformation associated to a rotation with angle $\theta\sim \epsilon$,
\eqa{
A_\perp^{L \prime}&=& A_\perp^{L} +  \epsilon A_\perp^{R*} \,, \nonumber \\
A_\|^{L \prime}&=& A_\|^{L} -  \epsilon A_\|^{R*} \,, \nonumber \\
A_0^{L \prime}&=& A_0^{L} -  \epsilon A_0^{R*} \,, \nonumber \\
A_\perp^{R \prime}&=& A_\perp^{R} -  \epsilon A_\perp^{L*}  - i \epsilon A_\perp^{R} k \,,\nonumber \\
A_\|^{R \prime}&=& A_\|^{R} +  \epsilon A_\|^{L*}  - i \epsilon A_\|^{R} k \,, \nonumber \\
A_0^{R \prime}&=& A_0^{R} +  \epsilon A_0^{L*}  - i \epsilon A_0^{R} k \,, \nonumber \\
|A_t^\prime|^2&=&|A_t|^2 - 2\epsilon \,[ \re (\azeL^2) - \re (\azeR^2) + k\, \im (\azeL^* \azeR )] \,,
\label{inf}}
where $k=[ (\re(A_\|^{L \, 2}) - \re(A_\|^{R \, 2}))- \left( \| \leftrightarrow \perp \right)]/[
\im (A_\|^L A_\|^{R*})+ \left( \| \leftrightarrow \perp \right)]$. Eq.~(\ref{inf}) together with  some important observations will guide us in the construction of the corresponding continuous transformation, namely:

\begin{itemize}
\item[I.] The structure of the distribution and the absence of lepton masses in the infinitessimal transformation informs us  that the symmetries of the massive distribution should be also symmetries of the massless case; consequently, the form of the rotation matrices should respect the form of the transformations in the massless case. 
\item[II.] The infinitessimal form shows that while the left components transform linearly,  the right components transform non-linearly. Moreover, in the limit of $k \to 0$,  the linear transformations of the massless case are recovered. 
\item[III.] All $J_i$, except for $J_{1s}$ and $J_{1c}$, are invariant under this infinitessimal transformation independently of the explicit form of $k$. This last remark is, indeed, connected to point I.
\end{itemize}
These considerations taken together imply that the continuous symmetry transformations $U_1(\theta)$ and $U_2(\tilde \theta)$ should take the form:
\eq{
  U_1(\theta)=
    \left[
    \begin{array}{rr}
      \cos \theta & -\sin \theta \\
     e^{-i \delta(\theta)} \sin \theta & e^{-i \delta(\theta)} \cos \theta
    \end{array} \right] \ , \quad 
  U_2(\tilde\theta)=
  \left[
    \begin{array}{rr}
      \cosh i \tilde{\theta} &  -\sinh i \tilde{\theta} \\
      -e^{-i \tilde\delta(\tilde\theta)} \sinh i \tilde{\theta} & e^{-i \tilde\delta(\tilde\theta)}\cosh i \tilde{\theta}
    \end{array}\right].
\label{symmass}}
The last step to determine these rotation matrices completely is to obtain the phases $\delta(\theta)$ and $\tilde\delta(\tilde\theta)$. These are non-linear functions of the transversity amplitudes and the angles $\theta$ and $\tilde\theta$, respectively. In a certain sense, the goal of these non-linear pieces  is to cure the breaking of the massless symmetry by the massive terms, while respecting the basic structure of the massless symmetry. Imposing that the mass term in $J_{1s}$
\eq{\re\left(\apeL\apeR^* + \apaL\apaR^*\right) \label{massterm}}
should be invariant under the symmetry transformation, we obtain that  $\sin\delta(\theta)$ and $\cos\delta(\theta)$ are just the result of a rotation of a unitary vector ${(\sin v, \cos v)}$ whose first component ($\sin v$) is indeed proportional to the mass term. This rotation is
\eq{
\binom{\sin\delta(\theta)}{\cos\delta(\theta)}=  \left(
    \begin{array}{rr}
      \cos u & -\sin u \\
      \sin u &  \cos u
    \end{array}\right)
\binom{\sin v}{\cos v} \,,
\label{sindelta}}
where $\sin v= x_1/\sqrt{h(\theta)}$, $\cos v=\eta_{y_1} \sqrt{1-x_1^2/h(\theta)}$  and  $\eta_{y_1}$  is the sign of the function $y_1$. The sign function  has been introduced to ensure that $\delta(0)=0$, which implies that  the transformation matrix becomes the identity matrix\footnote{Notice, however, that the $\cos v$ defined as $\cos v=-\eta_{y_1} \sqrt{1-x_1^2/h(\theta)}$ is also a solution, even if not connected to the identity in the limit $\theta \to 0$.} for $\theta\to 0$. The rotation ($u$) is defined by
\eq{
\cos u=\frac{y_1 \cos 2 \theta + y_2 \sin 2 \theta}{\sqrt{h(\theta)}}\,,  \quad \quad  \sin u=\frac{x_1 \cos 2 \theta + x_2 \sin 2 \theta}{\sqrt{h(\theta)}}\,,
\label{cosu}}
where
\eqa{  
\!\!\!\!\!&\!\!\!\!\!\!\!x_1\!\!\!&\!\!\!\!=\re (A_\|^L A_\|^{R*})+\left( \| \leftrightarrow \perp \right)
\,, \, \, \quad \;
x_2=\frac{1}{2}\left[ \re(A_\|^{L \, 2}) - \re(A_\|^{R \, 2}) \right]- \left( \| \leftrightarrow \perp \right) \,,
\nonumber \\
\!\!\!\!\!&\!\!\!\!\!\!\!y_1\!\!\!&\!\!\!\!=\im (A_\|^L A_\|^{R*})+ \left( \| \leftrightarrow \perp \right)
\,, \, \, \quad \;
y_2=\frac{1}{2}\left[ \im (A_\|^{L\, 2})  + \im (A_\|^{R \,2})\right] - \left( \| \leftrightarrow \perp \right)  \,,\nonumber \\
&h(\theta)&=(x_1 \cos 2 \theta + x_2 \sin 2 \theta)^2 + (y_1 \cos 2 \theta + y_2 \sin 2 \theta)^2 \,.
}
$x_1$ is precisely the mass term of $J_{1s}$. Being an invariant, it can be expressed in terms of the coefficients $J_i$ of the distribution:
\eq{x_1=\frac{q^2}{m_\ell^2} \frac{(2+\beta_\ell^2)}{4} \left( \frac{J_{1s}}{2+\beta_\ell^2} - \frac{J_{2s}}{\beta_\ell^2} \right).}
Also the non-linear parameter  $k$   in Eq.(\ref{inf}) can be rewritten as  $k=2 x_2/y_1$. Notice that due to the non-linear form of the transformation 
$$U_1(\theta_1) \cdot U_1(\theta_2) \not= U_1(\theta_1+\theta_2)\,.$$
This does not pose a problem since one can easily concatenate transformations, one after the other, always evaluating the corresponding $\delta(\theta_i)$ for each transformation. A final important remark is that the requirement of  $ \sqrt{1-x_1^2/h(\theta)}$ to be real imposes a restriction for the range of validity of the transformation around $\theta=0$, i.e., there is a maximum and a minimum allowed value for $\theta$, given by the condition
\eq{\left|\frac{x_1}{\sqrt{h(\theta)}}\right|\leq 1\,.
\label{condition}}
Inside this range of validity $\delta(\theta)$ is simply given by $\delta(\theta)=v-u$.
%
%

To complete the transformation of all amplitudes under $U_1(\theta)$ we still need to find the transformation of the amplitude $A_t$. This is  obtained by imposing the invariance of the mass term in $J_{1c}$
\eq{|A_t^\prime|^2= |A_t|^2 + 2 \left(
\re(\azeL^{}\azeR^{*}) - \re(\azeL^{\prime}\azeR^{* \prime}) \right)\,, \label{attransf} }
where the primed $A_0^{L,R}$ can be obtained easily from $n_0^\prime$. 

Following exactly the same steps for the symmetry transformation $U_2(\tilde\theta)$ we can also identify the corresponding $\tilde\delta(\tilde\theta)$ by the rotation
\eq{
\binom{\sin\tilde\delta(\tilde\theta)}{\cos\tilde\delta(\tilde\theta)}=  \left(
    \begin{array}{rr}
      \cos \tilde u & -\sin \tilde u \\
      \sin \tilde u &  \cos \tilde u
    \end{array}
  \right)
\binom{\sin\tilde v}{\cos\tilde v} \,,
\label{tildedelta}}
 where $\sin\tilde v= \tilde x_1/\sqrt{\tilde h(\tilde \theta)}$, $\cos \tilde v=\eta_{\tilde y_1} \sqrt{1-\tilde x_1^2/\tilde h(\tilde \theta)}$, $\eta_{\tilde y_1}$ is the sign of the function $\tilde y_1$  and
\eq{\cos \tilde u=\frac{\tilde y_1 \cos 2 \tilde \theta + \tilde y_2 \sin 2 \tilde \theta}{\sqrt{\tilde h(\tilde\theta)}}\,,  \quad  \quad  \sin \tilde u=\frac{\tilde x_1 \cos 2 \tilde \theta + \tilde x_2 \sin 2 \tilde \theta}{\sqrt{\tilde h(\tilde\theta)}}\ ,
\label{cosu}}
with
\eqa{ 
\!\!\!\!\!&\!\!\!\!\!\!\!\tilde x_1\!\!\!&\!\!\!\!=x_1 \,,
\, \, \, \,  \quad \;
\tilde x_2=\frac{1}{2}\left[ \im(A_\|^{L \, 2}) - \im(A_\|^{R \, 2}) \right]- \left( \| \leftrightarrow \perp \right) \,,
\nonumber \\
\!\!\!\!\!&\!\!\!\!\!\!\!\tilde y_1\!\!\!&\!\!\!\!=y_1 \,,
\, \, \, \, \quad \;
\tilde y_2=\frac{1}{2}\left[ -{\rm Re} (A_\|^{L\, 2})  - {\rm Re} (A_\|^{R \,2})\right] - \left( \| \leftrightarrow \perp \right)  \,, \nonumber \\
&\tilde h(\tilde\theta)&=(\tilde x_1 \cos 2 \tilde\theta + \tilde x_2 \sin 2 \tilde \theta)^2 + (\tilde y_1 \cos 2 \tilde\theta + \tilde y_2 \sin 2 \tilde \theta)^2 \,.
}
The same discussion about the range of validity can be applied to $U_2(\tilde\theta)$ just by substituting in Eq.~(\ref{condition}) $x_i \to \tilde x_i$, $h(\theta) \to \tilde h(\tilde\theta)$ and now $\tilde \delta(\tilde \theta)=\tilde v-\tilde u$.  $|A_t^\prime|^2$ can be computed from Eq. (\ref{attransf}) but using the corresponding primed $A_0^{L,R}$ amplitudes under the $U_2(\tilde\theta)$ transformation.

\subsection{Symmetries in the presence of scalar contributions}
\label{appscalar}

The last case to discuss here is the symmetry structure of the angular distribution in presence  of scalar contributions to the decay channel $B_d \to K^{*0} (\to K \pi) \ell^+ \ell^-$.

The infinitessimal transformation, too long to write it explicitly here, provides the following important information:
\begin{itemize}
\item $A_\perp$ and $A_\|$  transform exactly as in the massive case.
\item $A_0$ and $A_S$ get mixed in the transformation and, contrary to the massive case, an explicit dependence on the lepton mass appears in the symmetry transformation. This has an important consequence for the construction of observables: while the lepton mass terms in $J_{1s}$ is invariant by itself, the mass terms in 
$J_5$ and $J_7$ are not. 
\end{itemize}

Another fundamental difference with the previous cases, is that the use of the compact two-component complex vector  $n_i$  is no longer possible for $A_0$ and $A_S$. Consequently, we will introduce a new formalism in terms of four-component vectors 

\eq{\label{fourvectors}
\vec{v_\|}=\left(\begin{array}{c} \re\apaL \\ \im\apaL \\ \re\apaR \\ \im\apaR
\end{array} \right)  \ , \quad 
\vec{v_\perp}=\left(\begin{array}{c} \re\apeL \\ \im\apeL \\ -\re\apeR \\ -\im\apeR
\end{array} \right) \ , \quad
\vec{v_0}=\left(\begin{array}{c}  \re\azeL \\ \im\azeL \\ \re\azeR \\ \im\azeR \end{array} \right)
\ , \quad
\vec{v_S}=\left(\begin{array}{c}  \re A_S \\ \im A_S \\ 0 \\ 0 \end{array} \right)
}\\
and two $4\times 4$ matrices

\eq { \label{twomatrices}
C = \left (\begin {array} {cc} \mathbb{I}_2 & \mathbb{I}_2 \\ \mathbb{I}_2 & \mathbb{I}_2\end {array} \right)
\ ,\quad
\gamma=\left (\begin {array} {cc} - i \sigma_2 & 0 \\ 0  & i \sigma_2 \end {array} \right)\ 
}\\
and will describe the angular distribution in terms of them.
Notice that $\gamma^T=-\gamma$ and $\gamma.\gamma^T=\mathbb{I}_4$. We remind here the explicit form of the Pauli $\sigma$ matrices:

\eq{
\sigma_1 = \left (\begin {array} {cc} 0 & 1 \\ 1 & 0\end {array} \right)
\ ,\quad
\sigma_2 = \left (\begin {array} {cc} 0 & -i \\ i & 0\end {array} \right)
\ ,\quad
\sigma_3 = \left (\begin {array} {cc} 1 & 0 \\ 0 & -1\end {array} \right)
\ .
}\\
The matrix $C$ is needed to symmetrize the vector $\vec{v_S}$, which appears in the angular distribution as
 $\vec{v_{SC}}\equiv C\cdot\vec{v_S}$. In terms of the four vectors in Eq.~(\ref{fourvectors}) and the two matrices in Eq.(\ref{twomatrices}) one can rewrite the coefficients $J_i$ including scalars as

\eqa{
\hspace*{1cm} J_{1s}&=&\frac{(2 + \beta_\ell^2)}{4} (\vec{v_\perp}\cdot \vec{v_\perp} + \vec{v_\|}\cdot \vec{v_\|})+2 \frac{m_\ell^2}{q^2} (\vec{v_\|}\cdot (C-\mathbb{I}_4)\cdot \vec{v_\|} - \vec{v_\perp}\cdot (C-\mathbb{I}_4)\cdot \vec{v_\perp} )\,,
\nonumber \\
J_{1c}&=&\vec{v_0}\cdot \vec{v_0}+ 4 \frac{m_\ell^2}{q^2} (\vec{v_0}\cdot (C-\mathbb{I}_4)\cdot \vec{v_0}+|A_t|^2)+\frac{\beta_\ell^2}{2} (\vec{v_{SC}}\cdot \vec{v_{SC}})\,,
\nonumber\\
J_{2s}&=&\frac{\beta_\ell^2}{4} (\vec{v_\perp}\cdot \vec{v_\perp}+\vec{v_\|}\cdot \vec{v_\|})\,,
\quad \quad
J_{2c}=-\beta_\ell^2 (\vec{v_0}\cdot \vec{v_0})\,,
\nonumber\\
J_3&=&\frac{\beta_\ell^2}{2}(\vec{v_\perp}\cdot \vec{v_\perp}-\vec{v_\|}\cdot \vec{v_\|})\,,
\quad \quad
J_4=\frac{1}{\sqrt{2}}\beta_\ell^2 (\vec{v_\|}\cdot \vec{v_0})\,,
\nonumber
}
\eqa{
J_5&=&\sqrt{2} \beta_\ell (\vec{v_\perp}\cdot \vec{v_0}-\frac{m_\ell}{\sqrt{q^2}} \vec{v_\|}\cdot \vec{v_{SC}})\,,
\quad \quad
J_{6s}=2 \beta_\ell (\vec{v_\|}\cdot \vec{v_\perp})\,,
\nonumber \\
J_{6c}&=&4 \beta_\ell \frac{m_\ell}{\sqrt{q^2}} (\vec{v_0}\cdot  \vec{v_{SC}})\,,
\quad \quad
J_7=-\sqrt{2}\beta_\ell(\vec{v_\|}\cdot\gamma\cdot \vec{v_0}- \frac{m_\ell}{\sqrt{q^2}}\vec{v_\perp}\cdot\gamma\cdot \vec{v_{SC}})\,,
\nonumber \\
J_8&=&-\frac{1}{\sqrt{2}} \beta_\ell^2 (\vec{v_\perp}\cdot\gamma\cdot \vec{v_0})\,,
\quad \quad
J_9=-\beta_\ell^2 (\vec{v_\|}\cdot\gamma\cdot \vec{v_\perp})\,.
}

The angular distribution exhibits four symmetries:

\begin{itemize}

\item A global phase transformation\footnote{Incidentally notice that the matrix $\gamma$ can be interpreted also as a phase transformation of $\pi/2$ for the L components and -$\pi/2$ for the R components.} for
\eq{\vec{v_j^\prime}=V_{(0)}(\phi) \vec{v_j}\,,}
with $j=\|,\perp,0,SC$, and where
\begin{equation}
V_{(0)}=\left[\begin{array}{rrrr}
      \cos\phi &   -\sin\phi & 0 & 0 \\
      \sin\phi & \cos\phi & 0 & 0 \\
     0 & 0 & \cos\phi & -\sin\phi    \\
      0 & 0 & \sin\phi & \cos\phi     
        \end{array}\right]\,.
\end{equation}

\item An independent phase transformation for the $A_t$ amplitude: $A_t^\prime=e^{i \phi_t} A_t$ .

\item  And the same two rotations of the massive case. However, there is an important difference between the transformation properties of $A_\perp^{L,R}$, $A_\|^{L,R}$ on one side and the transformation properties of $A_0^{L,R}$, $A_S$ on the other, that will be detailed in the following.
\end{itemize}

On the one hand, $A_\perp^{L,R}$ and $A_\|^{L,R}$ transform exactly as in the massive case
\eq{\vec{v_j^\prime}=V_{(1)}(\theta) \vec{v_j} \quad {\rm and} \quad \vec{v_j^\prime}=V_{(2)}(\tilde\theta) \vec{v_j}}
for $j=\perp,\|$ and the matrices $V_{(1)}(\theta)$ and $V_{(2)}(\tilde\theta)$ are simply the mapping of the 2x2 matrices $U_1(\theta)$ and $U_2(\tilde\theta)$ in the 4-d formalism and are defined by
\begin{equation}
V_{(1)}(\theta)=\left[\begin{array}{rrrr}
      \cos\theta & 0 &  -\sin\theta & 0 \\
      0 & \cos\theta & 0 & \sin\theta \\
     \cos\delta \sin\theta & \sin\delta \sin\theta & \cos\delta \cos\theta & -\sin\delta \cos\theta     \\
      \sin\delta \sin\theta & -\cos\delta \sin\theta & \sin\delta \cos\theta & \cos\delta \cos\theta     
        \end{array}\right]
\end{equation}
and
\begin{equation}
 V_{(2)}(\tilde\theta)=\left[\begin{array}{rrrr}
      \cos\tilde\theta & 0 & 0 & -\sin\tilde\theta \\
      0 & \cos\tilde\theta & -\sin\tilde\theta & 0 \\
     -\sin\tilde\delta \sin\tilde\theta & \cos\tilde\delta \sin\tilde\theta & \cos\tilde\delta \cos\tilde\theta & -\sin\tilde\delta \cos\tilde\theta     \\
      \cos\tilde\delta \sin\tilde\theta & \sin\tilde\delta \sin\tilde\theta & \sin\tilde\delta \cos\tilde\theta & \cos\tilde\delta \cos\tilde\theta     
        \end{array}\right]\,.
\end{equation}\\
The non-linear structures $\delta(\theta)$ and $\tilde\delta(\tilde\theta)$ in $V_{(1)}(\theta)$ and $V_{(2)}(\tilde\theta)$  are the same as in Eq.~(\ref{sindelta}) and Eq.~(\ref{tildedelta}) respectively. In this new formalism, the expression of $x_i$, $y_i$, $\tilde x_i$ and $\tilde y_i$ can be rewritten in a more compact way:
\eqa{
x_1&=& \left(\vec{v_\|}\cdot (C-\mathbb{I}_4)\cdot \vec{v_\|}-\vec{v_\perp}\cdot
(C-\mathbb{I}_4)\cdot \vec{v}_\perp\right)/2 \,, \\
x_2&=& \left(\vec{v_\|}\cdot C_{x2}\cdot \vec{v_\|}-\vec{v_\perp}\cdot
C_{x2}\cdot \vec{v}_\perp\right)/2 \,, \\
y_1&=& \left(\vec{v_\|}\cdot C_{y1}\cdot \vec{v_\|}-\vec{v_\perp}\cdot
C_{y1}\cdot \vec{v}_\perp\right)/2 \,, \\
y_2&=&\left(\vec{v_\|}\cdot C_{y2}\cdot \vec{v_\|}-\vec{v_\perp}\cdot
C_{y2}\cdot \vec{v}_\perp\right)/2 
}
and
\eqa{
\tilde x_1&=& x_1\ ,\quad \quad \tilde y_1 = y_1 \,,  \\
\tilde x_2&=&\left(\vec{v_\|}\cdot C_{\tilde x2}\cdot \vec{v_\|}-\vec{v_\perp}\cdot
C_{\tilde x2}\cdot \vec{v}_\perp\right)/2 \,, \\
\tilde y_2&=& \left(\vec{v_\|}\cdot C_{\tilde y2}\cdot \vec{v_\|}-\vec{v_\perp}\cdot
C_{\tilde y2}\cdot \vec{v}_\perp\right)/2 \,,
}
with
\eqa{
C_{x2} &=& \left (\begin {array} {cc} \sigma_3 & 0 \\ 0
& -\sigma_3\end {array} \right)\ ,\quad
C_{y1} = \left (\begin {array} {cc} 0 & - i \sigma_2 \\ i \sigma_2
& 0\end {array} \right)\ ,\quad
C_{y2} =\left (\begin {array} {cc} \sigma_1 & 0 \\ 0
& \sigma_1\end {array} \right)\ ,\nn\\[3mm]
C_{\tilde x2} & = &\left (\begin {array} {cc} \sigma_1 & 0 \\ 0
& -\sigma_1\end {array} \right)\ ,\quad
C_{\tilde y2}  = \left (\begin {array} {cc} -\sigma_3 & 0 \\ 0
& -\sigma_3\end {array} \right)\ .
}\\
On the other hand, in the scalar case the transformation of the amplitudes $A_0^{L,R}$ is different from the massive case, but shares the same structure than the transformation of $A_S$:
\begin{equation}
\label{transf}
\vec{v_0^\prime}=V_{(j)} \left( { \hat{v_0}^{(j)}} + \vec{v_0} \right)\,,  \quad \quad   \vec{v_{SC}^\prime}=V_{(j)} \left( { \hat{v_S}^{(j)}} + \vec{v_{SC}} \right) \,,
\end{equation}
where 
\eq{ {\hat{v_0}^{(j)}}=\left(\begin{array}{c} { \hat{v_0}^{(j)}}_1 \\ { \hat{v_0}^{(j)}}_2 \\ { \hat{v_0}^{(j)}}_3\\ { \hat{v_0}^{(j)}}_4 \end{array} \right)  \ , \quad \quad
{ \hat{v_S}^{(j)}}=\left(\begin{array}{c}  { \hat{v_S}^{(j)}}_1\\ { \hat{v_S}^{(j)}}_2 \\ { \hat{v_S}^{(j)}}_3 \\ { \hat{v_S}^{(j)}}_4 \end{array} \right)\ ,}
with $j=1,2$. These vectors contain the non-linear part of the transformation associated to $A_0^{L,R}$ and $A_S$ and are functions of $\theta$ (for $j=1$), $\tilde\theta$ (for $j=2$), all amplitudes and the lepton mass $m_\ell$.

The last remaining point is to determine  ${ \hat{v_0}^{(j)}}$ and ${ \hat{v_S}^{(j)}}$. They  are fixed by requiring the invariance of the angular distribution under the $V_1(\theta)$ and $V_2(\tilde\theta)$ transformations.  
The set of eight equations required to obtain the components of ${ \hat{v_0}^{(j)}}$ and ${ \hat{v_S}^{(j)}}$ arise from the coefficients $J_4$, $J_5$, $J_7$ and $J_8$ after imposing the invariance of the angular distribution under the transformations:\\[-4mm]
\begin{eqnarray}
&\vec{v_\|} \cdot \hat{v_0}^{(j)}=0 \ , \quad \quad
\vec{v_\perp} \cdot \gamma \cdot \hat{v_0}^{(j)}=0\ ,& \nonumber \\
&\vec{v_\perp} \cdot  \hat{v_0}^{(j)} - \frac{m_l}{\sqrt{q^2}} \vec{v_\|} \cdot \hat{v_S}^{(j)}=0 \ , \quad \quad
\vec{v_\|} \cdot \gamma \cdot \hat{v_0}^{(j)} - \frac{m_l}{\sqrt{q^2}} \vec{v_\perp} \cdot \gamma \cdot \hat{v_S}^{(j)}=0\ ,&
\label{eqsv0}\end{eqnarray}\\
together with the ones derived from $J_{2c}$ and $J_{6c}$

\begin{equation} \label{lasttwo}
\hat{v_0}^{(j)} \cdot \hat{v_0}^{(j)} + 2 \hat{v_0}^{(j)} \cdot \vec{v_0}=0 \ , \quad \quad
\hat{v_0}^{(j)} \cdot \hat{v_S}^{(j)} + \hat{v_0}^{(j)} \cdot \vec{v_{SC}}+ \vec{v_0} \cdot \hat{v_S}^{(j)}=0\ ,
\end{equation}\\
and  two more equations to impose that the first component of the vector  $\vec{v_{SC}}^\prime$ must equal the third while the second component should equal the fourth\footnote{Therefore 
$ \vec{v_{SC}}^\prime=({\re A_S^\prime,\im A_S^\prime,\re A_S^\prime,\im A_S^\prime})$.}

\begin{equation}
\label{dsI}
\vec{d}^{(j)} \cdot ( \hat{v_S}^{(j)}+  \vec{v_{SC}} )=0 \ , \quad \quad
\vec{e}^{(j)} \cdot ( \hat{v_S}^{(j)}+  \vec{v_{SC}} )=0\ ,
\end{equation}
where
\begin{eqnarray} 
\vec{d}^{(j)}&=& ( V_{(j) \, 1,1}-V_{(j) \, 3,1},  V_{(j) \, 1,2}-V_{(j) \, 3,2},  V_{(j) \, 1,3}-V_{(j) \, 3,3},  V_{(j) \, 1,4}-V_{(j) \, 3,4} )\ ,\nonumber \\
\vec{e}^{(j)}&=&( V_{(j) \, 2,1}-V_{(j) \, 4,1},  V_{(j) \, 2,2}-V_{(j) \, 4,2},  V_{(j) \, 2,3}-V_{(j) \, 4,3},  V_{(j) \, 2,4}-V_{(j) \, 4,4} )\ .
\end{eqnarray}\\[-3mm]
The  system can be solved as follows:
\begin{itemize}

\item[{I.}] Obtain $\hat{v_0}^{(j)}$ in terms of  $\hat{v_S}^{(j)}$ from the first set of Eqs.(\ref{eqsv0}). The components of the vector $\hat{v_0}^{(j)}$ are then:\\[-5mm]
\begin{eqnarray}\label{v01234}
\!\!\!\!\!{ \hat{v_0}^{(j)}}_1&=& m_1^{(j)} \re X^{L,R} + m_2^{(j)} \im Y^{L,R}\,, \quad { \hat{v_0}^{(j)}}_2= m_1^{(j)} \im X^{L,R} - m_2^{(j)} \re Y^{L,R}	\ , \nonumber \\
\!\!\!\!\!{ \hat{v_0}^{(j)}}_3&=& -m_1^{(j)} \re X^{R,L} - m_2^{(j)} \im Y^{R,L}\,, \quad { \hat{v_0}^{(j)}}_4= -m_1^{(j)} \im X^{R,L} +m_2^{(j)} \re Y^{R,L}\,, \quad
\end{eqnarray}\\[-3mm]
where
\begin{eqnarray}
X^{a,b}&=& A_\|^a A_\perp^{*b} A_\|^b +  |A_\|^b|^2 A_\perp^a\ ,
\nonumber \\
Y^{a,b}&=& A_\perp^a A_\|^{*b} A_\perp^b +  |A_\perp^b|^2 A_\|^a\ ,
\end{eqnarray}
with\\[-5mm]
\begin{equation}
\label{mij}
m_1^{(j)}=\frac{1}{\omega}\frac{m_\ell}{\sqrt{q^2}} \vec{v_\|} \cdot \hat{v_S}^{(j)}  \ , \quad
m_2^{(j)}=\frac{1}{\omega}\frac{m_\ell}{\sqrt{q^2}} \vec{v_\perp}\cdot\gamma\cdot\hat{v_S}^{(j)}
\end{equation}
and $\omega=|\apeL|^2|\apaR|^2+|\apeR|^2|\apaL|^2 + 2 \re(\apaL^*\apeL\apeR\apaR^*)$.\\

\item[{II.}] Use Eq.(\ref{dsI}) to  express the components $\hat{v_S}^{(j)}_3$ and $\hat{v_S}^{(j)}_4$ in terms of $\hat{v_S}^{(j)}_1$ and $\hat{v_S}^{(j)}_2$:
\begin{eqnarray}
\hat{v_S}^{(j)}_3&=&\frac{d_1^{(j)} e_4^{(j)} - d_4^{(j)} e_1^{(j)}}{d_4^{(j)}  e_3^{(j)}  - d_3^{(j)}  e_4^{(j)} } \left[\hat{v_S}^{(j)}_1+ \re A_S \right] +
\frac{d_2^{(j)}  e_4^{(j)}  - d_4^{(j)}  e_2^{(j)} }{d_4^{(j)}  e_3^{(j)}  - d_3^{(j)}  e_4^{(j)} } \left[\hat{v_S}^{(j)}_2+\im A_S \right] - \re A_S \nn \\
\hat{v_S}^{(j)}_4&=&\frac{d_3^{(j)} e_1^{(j)} - d_1^{(j)} e_3^{(j)}}{d_4^{(j)}  e_3^{(j)}  - d_3^{(j)}  e_4^{(j)} } \left[\hat{v_S}^{(j)}_1+ \re A_S \right] +
\frac{d_3^{(j)}  e_2^{(j)}  - d_2^{(j)}  e_3^{(j)} }{d_4^{(j)}  e_3^{(j)}  - d_3^{(j)}  e_4^{(j)} } \left[\hat{v_S}^{(j)}_2+\im A_S \right] - \im A_S\nn \\
\label{vs34}
\end{eqnarray}
\item[{III.}] Determine $\hat{v_S}^{(j)}_1$ and $\hat{v_S}^{(j)}_2$. Using  Eqs.(\ref{lasttwo}) together with Eq.(\ref{v01234}) we find the last two equations. The first of Eqs.~(\ref{lasttwo}) gives
\begin{eqnarray} \label{numer}
&[|X^{L,R}|^2 m_1^{(j) \, 2} + |Y^{L,R}|^2 m_2^{(j) \, 2} + 2 m_1^{(j)} m_2^{(j)} \im (X^{L,R *} Y^{L,R})]+(L \leftrightarrow R)=& \nonumber \\ &[- 2 \re (A_0^{L*} X^{L,R}) m_1{(j)} - 2 \im (A_0^{L*} Y^{L,R}) m_2^{(j)}] - (L \leftrightarrow R)&
\end{eqnarray} 
where $m_i^{(j)}$ are functions of  $\hat{v_S}^{(j)}$ as given in Eq.(\ref{mij}). 

The second of Eqs.(\ref{lasttwo}) gives rise to
\begin{eqnarray}
[ m_1^{(j)} \re (A_S^* X^{L,R})\, +&\!\!\!\!\!\!\!\!\!\!\!\!\!\!\!\!\!\!\!\!\!\!\!\!\!\!\!\!\!\!\!\!m_2^{(j)} \im (A_S^* Y_{L,R})] - (L \leftrightarrow R) =
\label{numer2} \\
& -\hat{v_S}^{(j)}_1 (\re A_0^L + m_1^{(j)} \re X^{L,R} + m_2^{(j)} \im Y^{L,R} ) \nonumber \\
& - \hat{v_S}^{(j)}_2 (\im A_0^L + m_1^{(j)} \im X^{L,R} - m_2^{(j)} \re Y^{L,R} ) 
\nonumber \\ 
& -\hat{v_S}^{(j)}_3 (\re A_0^R - m_1^{(j)} \re X^{R,L} - m_2^{(j)} \im Y^{R,L} ) \nonumber \\
&- \hat{v_S}^{(j)}_4 (\im A_0^R - m_1^{(j)} \im X^{R,L} + m_2^{(j)} \re Y^{R,L} )\,. \nonumber
 \end{eqnarray}
\item[IV.] Substituting Eqs.~(\ref{mij}) and  Eqs.~(\ref{vs34}) in Eq.~(\ref{numer}) and Eq.~(\ref{numer2}) we end up with  a system of two coupled quadratic equations which are function of ${ \hat{v_S}^{(j)}}_1$ and ${ \hat{v_S}^{(j)}}_2$. This system  can be solved numerically and typically provides  two complex solutions (to be discarded) and two real ones. The real solutions for  ${ \hat{v_S}^{(j)}}_1$ and ${ \hat{v_S}^{(j)}}_2$, once inserted in Eq.~(\ref{transf}), generate two sets of transformed amplitudes $\vec{v_0^\prime}$ and $\vec{v_{SC}^{\prime}}$ that leave the angular distribution invariant.  One of them is connected to the identity whereas the other is not, exactly as it occurred with $\delta(\theta)$ and $\tilde\delta(\tilde\theta)$.  
This completes the definition of the symmetry transformation of the $A_0^{L,R}$ and $A_S$ amplitudes. 

\end{itemize}

\section{Large recoil limit expressions}
\label{largerec}

In this appendix we present the expressions of the observables $P_i$ and $M_i$ in the large recoil limit. These are useful to study qualitative properties of the observables as well as for rough quantitative estimates.
\begin{eqnarray}
|n_0|^2 &=& 2 Q_1^2 Q_2^2 \xi_\|^2 \left(|\C{10}-\Cp{10}|^2+|(F \hat{s}\ \Ceff{7}+\Ceff{9})-(F \hat{s}\ \Cpeff{7}+\Cpeff{9})|^2\right)\,, \\[2mm]
|n_\||^2 &=& 2 Q_1^2 \xi_\bot^2 \left(|\C{10}-\Cp{10}|^2+|(F \Ceff{7}+\Ceff{9})-(F \Cpeff{7}+\Cpeff{9})|^2\right)\,, \\[2mm]
|n_\bot|^2 &=& 2 Q_1^2 \xi_\bot^2 \left(|\C{10}+\Cp{10}|^2+|(F \Ceff{7}+\Ceff{9})+(F \Cpeff{7}+\Cpeff{9})|^2\right)\,, \\[2mm]
{\rm{Re}}(n_\bot^\dagger n_\|) &=& 4 Q_1^2 \xi_\bot^2 {\rm{Re}}[( F \Ceff{7}+\Ceff{9}) \C{10}^*- (F \Cpeff{7}+\Cpeff{9}) \C{10}^{\prime*}]\,, \\[2mm]
{\rm{Im}}(n_\bot^\dagger n_\|) &=& -4 Q_1^2 \xi_\bot^2 {\rm{Im}}[\C{10}\C{10}^{\prime*}+(F \Ceff{7}+\Ceff{9})(F \Cpeff{7}+\Cpeff{9})^*]\,, \\[2mm]
{\rm{Re}}(n_0^\dagger n_\|) &=& 2 Q_1^2 Q_2 \xi_\| \xi_\bot \left(|\C{10}-\Cp{10}|^2+|\Ceff{9}-\Cpeff{9}|^2+F^2 \hat{s}|\Ceff{7}-\Cpeff{7}|^2\right. \nn \\
 &&\left.+F(1+\hat{s}){\rm{Re}}[(\Ceff{7}-\Cpeff{7})(\Ceff{9}-\Cpeff{9})^*]\right) \,,\\[2mm]
{\rm{Im}}(n_0^\dagger n_\|) &=& -2 Q_1^2 Q_2 \xi_\| \xi_\bot  {\rm{Im}}[F (1-\hat{s}) (\Ceff{7}-\Cpeff{7})(\C{10}-\Cp{10})^*]  \,, \\[2mm]
{\rm{Re}}(n_0^\dagger n_\bot) &=& 2 Q_1^2 Q_2 \xi_\| \xi_\bot {\rm{Re}}[((F (1+\hat{s}) \Ceff{7} + \Ceff{9})+(F (1-\hat{s}) \Cpeff{7}+\Ceff{9}))\C{10}^* \nn \\
 &&-((F (1-\hat{s}) \Ceff{7}+\Cpeff{9})+(F (1+\hat{s}) \Cpeff{7}+ \Cpeff{9})) \C{10}^{\prime*}] \,,\\[2mm]
{\rm{Im}}(n_0^\dagger n_\bot) &=& 2 Q_1^2 Q_2 \xi_\| \xi_\bot \left(2{\rm{Im}}[\C{10} \C{10}^{\prime*}+F^2 \hat{s}\ \Ceff{7} {\Cpeff{7}}^{*}+ \Ceff{9} {\Cpeff{9}}^{*}] -{\rm{Im}}[F ((1-\hat{s}) \Ceff{7} \right.\nn  \\ 
&&\left.+(1+\hat{s}) \Cpeff{7}) {\Ceff{9}}^*-F ((1+\hat{s}) \Ceff{7}+(1-\hat{s}) \Cpeff{7}) {\Cpeff{9}}^{*}]\right)\,,\\[2mm]
|n_\bot|^2 - |n_\||^2 &=& 8 Q_1^2 \xi_\bot^2 \left({\rm{Re}}[ \C{10} \C{10}^{\prime*}]+{\rm{Re}}[(F \Ceff{7}+\Ceff{9})(F \Cpeff{7}+\Cpeff{9})^*]\right)\,,\\[2mm]
|n_\||^2 + |n_\bot|^2&=& 4 Q_1^2 \xi_\bot^2 \left(|\C{10}|^2+|\Cp{10}|^2+|F \Ceff{7}+\Ceff{9}|^2+|F \Cpeff{7}+\Cpeff{9}|^2\right) \,, \\[2mm]
M_1 &=&  - \frac{2 m_\ell^2}{q^2 \beta_\ell^2}\frac{|\C{10}|^2+|\Cp{10}|^2-|F \Ceff{7}+\Ceff{9}|^2-|F \Cpeff{7}+\Cpeff{9}|^2}{|\C{10}|^2+|\Cp{10}|^2+|F \Ceff{7}+\Ceff{9}|^2+|F \Cpeff{7}+\Ceff{9}|^2}\,, \\[2mm]
M_2  &=&  \frac{4 m_\ell^2}{q^2}.
\end{eqnarray}
where the following short-hand notation has been used
\begin{equation}
F \equiv \frac{2 \hat{m}_b}{ \hat{s}}, \quad 
Q_1 \equiv \sqrt{2} N m_B (1-\hat{s}), \qquad
Q_2 \equiv \frac{1}{2 \sqrt{2} \hat{m}_{K^*} \sqrt{\hat{s}}} (1-\hat{s})\ .
\end{equation}



\end{document}